%% file: main_arXiv_ver_2.tex
\documentclass[11pt,a4paper]{article}

\input{no-revtex}

\input{jm-tex-macros-public}
\usepackage[utf8]{inputenc}
\usepackage{comment}
\usepackage{float}
\usepackage{xcolor}
\usepackage{physics}
\usepackage{array}
\usepackage{ragged2e}
\usepackage{booktabs}
\usepackage{latexsym}
\usepackage{mathtools}

\usepackage{subcaption}
\usepackage{graphicx}
\usepackage{subcaption}
\usepackage{cleveref}
\usepackage{amsfonts,amsmath,bm}
\usepackage[toc]{appendix}
\usepackage{subfloat}

\numberwithin{equation}{section}
\definecolor{darkgreen}{rgb}{0,0.5,0}
\definecolor{darkblue}{rgb}{0,0,0.6}

\hypersetup{
	citecolor=blue,
	colorlinks=true,
	filecolor=red,
	linkcolor=blue,
	linktocpage=true,
	urlcolor=darkgreen
}

\newcount\hour
\newcount\minute
\newtoks\amorpm
\hour=\time\divide\hour by60
\minute=\time{\multiply\hour by60 \global\advance\minute by-\hour}
\edef\standardtime{{\ifnum\hour<12 \global\amorpm={am}%
		\else\global\amorpm={pm}\advance\hour by-12 \fi
		\ifnum\hour=0 \hour=12 \fi
		\number\hour:\ifnum\minute<10 0\fi\number\minute\the\amorpm}}
\edef\militarytime{\number\hour:\ifnum\minute<10 0\fi\number\minute}

\def\draftlabel#1{{\@bsphack\if@filesw {\let\thepage\relax
			\xdef\@gtempa{\write\@auxout{\string
					\newlabel{#1}{{\@currentlabel}{\thepage}}}}}\@gtempa
		\if@nobreak \ifvmode\nobreak\fi\fi\fi\@esphack}
	\gdef\@eqnlabel{#1}}
\def\@eqnlabel{}
\def\@vacuum{}
\def\draftmarginnote#1{\marginpar{\raggedright\scriptsize\tt#1}}

\def\draft{\oddsidemargin -.5truein
		\def\@oddfoot{\sl preliminary draft \hfil
			\rm\thepage\hfil\sl\today\quad\militarytime}
		\let\@evenfoot\@oddfoot \overfullrule 3pt
		\let\label=\draftlabel
		\let\marginnote=\draftmarginnote
		\def\@eqnnum{(\theequation)\rlap{\kern\marginparsep\tt\@eqnlabel}%
			\global\let\@eqnlabel\@vacuum}  }
	
	\def\preprint{\twocolumn\sloppy\flushbottom\parindent 2em
		\leftmargini 2em\leftmarginv .5em\leftmarginvi .5em
		\oddsidemargin -.5in    \evensidemargin -.5in
		\columnsep .4in \footheight 0pt
		\textwidth 10.in        \topmargin  -.4in
		\headheight 12pt \topskip .4in
		\textheight 6.9in \footskip 0pt
		\def\@oddhead{\thepage\hfil\addtocounter{page}{1}\thepage}
		\let\@evenhead\@oddhead \def\@oddfoot{} \def\@evenfoot{} }
	
	\def\numberbysection{\@addtoreset{equation}{section}
		\def\theequation{\thesection.\arabic{equation}}}
	
	\def\underline#1{\relax\ifmmode\@@underline#1\else
		$\@@underline{\hbox{#1}}$\relax\fi}

	\def\titlepage{\@restonecolfalse\if@twocolumn\@restonecoltrue\onecolumn
		\else \newpage \fi \thispagestyle{empty}\c@page\z@
		\def\thefootnote{\fnsymbol{footnote}} }
	
	\def\endtitlepage{\if@restonecol\twocolumn \else \newpage \fi
		\def\thefootnote{\arabic{footnote}}
		\setcounter{footnote}{0}}  

\catcode`@=12
\relax

\def\figcap{\section*{Figure Captions\markboth
		{FIGURECAPTIONS}{FIGURECAPTIONS}}\list
	{Figure \arabic{enumi}:\hfill}{\settowidth\labelwidth{Figure
			999:}
		\leftmargin\labelwidth
		\advance\leftmargin\labelsep\usecounter{enumi}}}
 \relax
\def\tablecap{\section*{Table Captions\markboth
		{TABLECAPTIONS}{TABLECAPTIONS}}\list
	{Table \arabic{enumi}:\hfill}{\settowidth\labelwidth{Table
			999:}
		\leftmargin\labelwidth
		\advance\leftmargin\labelsep\usecounter{enumi}}}
 \relax
\def\reflist{\section*{References\markboth
		{REFLIST}{REFLIST}}\list
	{[\arabic{enumi}]\hfill}{\settowidth\labelwidth{[999]}
		\leftmargin\labelwidth
		\advance\leftmargin\labelsep\usecounter{enumi}}}
 \relax
\makeatletter
\newcounter{pubctr}
\def\publist{\@ifnextchar[{\@publist}{\@@publist}}
\def\@publist[#1]{\list
	{[\arabic{pubctr}]\hfill}{\settowidth\labelwidth{[999]}
		\leftmargin\labelwidth
		\advance\leftmargin\labelsep
		\@nmbrlisttrue\def\@listctr{pubctr}
		\setcounter{pubctr}{#1}\addtocounter{pubctr}{-1}}}
\def\@@publist{\list
	{[\arabic{pubctr}]\hfill}{\settowidth\labelwidth{[999]}
		\leftmargin\labelwidth
		\advance\leftmargin\labelsep
		\@nmbrlisttrue\def\@listctr{pubctr}}}
 \relax
\makeatother
\newskip\humongous \humongous=0pt plus 1000pt minus 1000pt

\newif\ifdtup

\relax

\def\be{\begin{equation}}
	\def\ee{\end{equation}}
\def\ba{\begin{eqnarray}}
	\def\ea{\end{eqnarray}}

\def\e{\epsilon}

\def\m{\mu}

\def\n{\nu}

\def\S{\Sigma}

\def\IR{\relax{\rm I\kern-.18em R}}

\def\pp{\partial}

\def\IR{\relax{\rm I\kern-.18em R}}
\def\IL{\relax{\rm I\kern-.18em L}}

\def\inv{^{\raise.15ex\hbox{${\scriptscriptstyle -}$}\kern-.05em 1}}

\def\bea{\begin{eqnarray}}
	\def\eea{\end{eqnarray}}

\def\e{\epsilon}

\def\m{\mu} \def\n{\nu}

  \def\S{\Sigma}
\def\t{\tau}

\definecolor{markcolor2}{rgb}{1,0,0}

\definecolor{markcolor3}{rgb}{0,1,0}

\begin{document}
	
	\thispagestyle{empty}

	\title{\Large\bf\boldmath Aspects of holographic timelike entanglement entropy \\ in black hole backgrounds}

\vspace{0pt}	

\author{
Mir Afrasiar${}^{1}$,
Jaydeep Kumar Basak${}^{2}$ and Keun-Young Kim${}^{2,3}$

}
\address{
{\it ${}^1$
	Department of Physics, Shanghai University, 99 Shangda Road, Shanghai, 200444, China\\}
{\it ${}^2$
	Department of Physics and Photon Science, Gwangju Institute of Science and Technology, \\
	123 Cheomdan-gwagiro, Gwangju 61005, Korea\\}
{\it ${}^3$
	Research Center for Photon Science Technology, Gwangju Institute of Science and Technology, \\
123 Cheomdan-gwagiro, Gwangju 61005, Korea\\}
}

\date{}			

\email{\href{mailto:mirhepth@gmail.com}{mirhepth@gmail.com}, \href{mailto:jkb.hep@gmail.com}{jkb.hep@gmail.com},
\href{mailto:fortoe@gist.ac.kr}{fortoe@gist.ac.kr}
}

\abstract{
We study the holographic construction of timelike entanglement entropy (tEE) in black hole backgrounds in Lorentzian geometries. The holographic tEE is realized through extremal surfaces consisting of spacelike and timelike branches that encode its real and imaginary components, respectively. In the BTZ black hole, these surfaces extend into the interior of the black hole and reproduce the field-theoretic results. The analysis is further generalized to higher-dimensional AdS-Schwarzschild black holes, where the characteristics of tEE are obtained with increasing size of the boundary subsystem. Besides, we also show that the boundary subsystem length diverges at a dimension-dependent critical turning point. Notably, this critical point moves closer to the black hole horizon as the dimensionality of the bulk increases. For large subsystem lengths, the finite part of the tEE displays a characteristic volume-plus-area structure, with a real volume term and a complex coefficient of the area term approaching constant values at large dimensions. Besides, we also study the monotonicity of a new quantity, timelike entanglement density, which offers insights into a timelike area theorem in specific limits. Subsequently, we investigate the near-horizon dynamics in various black hole backgrounds, where the spacelike and timelike surfaces exhibit exponential growth of the form $e^{\frac{2\pi}{\beta} \Delta t}$ with inverse black hole temperature $\beta$.

}

\vspace{1.15in}

\begin{center}
{\it Dedicated to the memories of Tamal Krishna Bhattacharya (Tamal Da) and Prahlad Gupta (Prahlad Da)}
\end{center}

\newpage
\tableofcontents

\newpage

\section{Introduction}\label{sec_intro}

Entanglement has emerged as a central concept bridging diverse areas of theoretical and experimental research, ranging from quantum information and many-body physics to high-energy physics. Among its many characterizations, an entanglement measure, entanglement entropy, has played a particularly significant role, providing a striking connection between field-theoretic computations and geometric quantities in gravitational physics through the celebrated AdS/CFT correspondence \cite{Ryu:2006bv,Ryu:2006ef}. Despite its success, the definition and computation of entanglement entropy inherently rely on a choice of time slice, thereby restricting its applicability to constant-time settings. As a result, entanglement entropy becomes inadequate for probing the dynamical growth of quantum correlations in time-dependent systems.

This limitation has been addressed in recent years through the introduction of pseudoentropy \cite{Mollabashi:2020yie,Mollabashi:2021xsd}, which generalizes the notion of entanglement entropy. As a novel extension, pseudoentropy reduces to entanglement entropy in an appropriate limit, a feature that has attracted considerable attention in the scientific community. At its core, the definition of pseudoentropy also relies on a bipartition of the Hilbert space of a quantum system, $\mathcal{H}=\mathcal{H_A}\otimes\mathcal{H}_{A^c}$ but in contrast to entanglement entropy, it employs the transition matrix $\t=|\psi\rangle\langle \phi|/\langle\phi| \psi\rangle$ where $|\psi\rangle$ and $|\phi\rangle$ denote two generic pure states, corresponding to different quantum systems or to the same system at different times. In the special case where the two states coincide, the transition matrix reduces to the standard reduced density matrix used in the definition of entanglement entropy. Analogous to entanglement entropy, the pseudoentropy of a subsystem $A$ is defined as $S_A=-\text{Tr}\left[\tau_A\log\tau_A\right]$ where $\tau_A$ is the reduced transition matrix. Since $\tau$ is not necessarily Hermitian, the pseudoentropy can in general be complex. Interestingly, while the real part shows sensitivity to chaotic dynamics and phase transitions \cite{He:2024jog,Goto:2021kln,Kanda:2023jyi,Chen:2025ibe,Shinmyo:2023eci}, the interpretation of the imaginary part remains less understood, and continues to be an active area of research \cite{Caputa:2024gve}. A particularly intriguing context arises in the dS/CFT correspondence, where reduced density matrices in the dual Euclidean CFT are inherently non-Hermitian. In this setting, pseudoentropy naturally emerges as the appropriate measure, as demonstrated in \cite{Doi:2022iyj}, where a novel relation was unravelled between pseudoentropy in dS/CFT and a time-like generalization of entanglement entropy in AdS/CFT. This new quantity, termed the timelike entanglement entropy (tEE), will be the central focus of our study. Conceptually, tEE can be viewed as a natural extension of entanglement entropy along the time direction, thereby offering a new perspective on causal structure and temporal correlations in quantum systems.

The central developments of timelike entanglement entropy (tEE) arise in holographic scenarios, where this quantity provides new insights into the emergence of time \cite{Doi:2023zaf}.\footnote{A connection between the properties of timelike entanglement entropy and the non-Hermitian transition matrix (also called as generalized density matrix) was explored in \cite{Milekhin:2025ycm,Guo:2025dtq}. Subsequently, the authors of \cite{Das:2025fcd} extended the construction of the generalized density matrix involving a two-point correlator to a four-point (generalized to $2N$ points) correlator, which nicely captures various chaotic properties of a quantum many-body system. Recently, in \cite{Harper:2025lav}, timelike entanglement entropy, its holographic duals and the degree of non-Hermiticity of a given density matrix (utilizing a measure called imagitivity) were studied in detail for different classes of non-Hermitian density matrices.} In \cite{Doi:2023zaf}, the holographic dual of tEE was proposed as a combination of spacelike and timelike surfaces homologous to the timelike subsystem defined in the boundary CFT. An alternative approach considered in the same work involved a unique surface in the Euclidean background, similar to the Ryu–Takayanagi surface, for a temporal subsystem in the boundary, followed by a Wick rotation of the time coordinate to obtain the tEE. Note that in \cite{Grieninger:2023knz}, the Euclidean surface prior to Wick rotation was introduced as a distinct quantity termed the temporal entanglement entropy, which was extensively studied in the context of renormalization group flows. For tEE, both the holographic approaches were shown to coincide in the context of AdS$_3$/CFT$_2$ as well as in certain higher-dimensional setups, such as hyperbolic subsystems.\footnote{See \cite{Narayan:2022afv,Li:2022tsv,Jena:2024tly,Jiang:2023ffu,Narayan:2023ebn,Jiang:2023loq,Das:2023yyl,Chu:2023zah,Guo:2024lrr,Basu:2024bal,Anegawa:2024kdj,Heller:2024whi,Chang:2024voo,Xu:2024yvf,Wen:2024yny,Roychowdhury:2025ukl,Jiang:2025pen,Katoch:2025bnh,Roychowdhury:2025aye,Chu:2025sjv,Nunez:2025gxq,Nunez:2025ppd,Heller:2025kvp,Roychowdhury:2025ebs,Gong:2025pnu,Nunez:2025puk,Zhao:2025zgm,Nanda:2025tid,Guo:2025pru} and the references therein, for various recent developments involving timelike entanglement entropy.} Note that, unlike the approaches in \cite{Doi:2022iyj,Doi:2023zaf}, different studies have proposed a holographic construction of tEE involving bulk surfaces in complexified spacetime coordinates \cite{Heller:2024whi,Heller:2025kvp,Nunez:2025ppd,Nunez:2025puk} which also match with the results obtained by Wick rotation. However, in more general quantum field theories with reduced symmetry, i.e., non-conformal and non-relativistic theories, the Wick-rotation-based definition becomes technically challenging. Building on this idea, the authors of \cite{Basak:2023otu,Afrasiar:2024lsi,Afrasiar:2024ldn} proposed a modified holographic prescription of the timelike entanglement entropy from the equation of motion. Utilizing the freedom to choose the signature of the integration constant in the equation of motion, two different surfaces were observed as candidate solutions. A positive integration constant corresponds to a set of two space-like surfaces starting from the endpoints of the subsystem at the boundary and expanding into the bulk. The negative value of the constant indicates a timelike surface with a turning point in the bulk extending from the turning point to the interior of the bulk. The areas of the spacelike and the timelike surfaces were computed to be real and imaginary, respectively. In AdS$_3$/CFT$_2$ scenario with zero temperature, this holographic computation matches exactly with the field theoretic result of tEE.
In non-conformal theories, a robust merging condition allows the timelike and the spacelike surfaces to be joined rigorously into a single continuous surface homologous to the boundary temporal subsystem, providing a unique geometric dual for tEE. Importantly, in such theories, the holographic tEE is also shown to be sensitive to phase transitions. In higher-dimensional setups with hyperscaling violation and Lifshitz-like anisotropy, the holographic analysis of tEE revealed that both the spacelike and timelike surfaces encode important aspects of the stability and naturalness theory.
Besides, the approach in \cite{Afrasiar:2024lsi,Afrasiar:2024ldn} offers a technique to read the size of the temporal subsystem from a holographic computation, which will be crucial in the context of the present article. Utilizing this construction, a holographic timelike $c$-function was proposed which captures the degrees of freedom along the renormalization group (RG) flow in an anisotropic theory with null energy conditions and thermodynamic stability \cite{Giataganas:2025div,Giataganas:2025ize}.

In this article, we investigate the holographic timelike entanglement entropy (tEE) in black hole backgrounds. We begin with the simplest case of BTZ black holes, following the holographic construction discussed in \cite{Afrasiar:2024lsi,Afrasiar:2024ldn}. The analysis in \cite{Doi:2023zaf} suggests that the holographic tEE consists of a timelike surface and a pair of spacelike surfaces, where the entire timelike surface together with part of the spacelike surfaces remain confined to the interior of the black hole. In contrast, our results demonstrate that the equations of motion allow both spacelike and timelike surfaces to extend from the interior to the exterior region of the black hole. Despite their different geometric structures, the holographic tEE obtained in this work matches exactly with the CFT and the bulk results in \cite{Doi:2023zaf}. Subsequently, we extend our analysis to higher-dimensional Schwarzschild black holes, where an analogous geometric structure of spacelike and timelike extremal surfaces arises. In these backgrounds, we identify the existence of a critical turning point associated with the timelike surface, which corresponds to an infinitely long subsystem at the asymptotic boundary. Notably, this critical point moves closer to the black hole horizon as the spacetime dimension increases. Subsequently, the areas of the spacelike and timelike extremal surfaces are obtained numerically with specific dimensions $d\geq 3$ as functions of the subsystem size. We renormalize the tEE by subtracting the contribution of the disconnected surfaces with vanishing slope, which extend from the asymptotic boundary to the horizon. This procedure removes only the divergent contribution to the real part of the tEE, leaving its imaginary component unchanged. Furthermore, we analyze how the areas of the spacelike and timelike surfaces depend on the subsystem size, which is holographically encoded in the turning point $r_{0}$ of the timelike surface. We further examine the UV-finite holographic tEE in the higher-dimensional black hole and show that it naturally decomposes into a sum of a volume term and an area term when the boundary subsystem length is considered to be very long. Our results indicate that the volume term is purely real, while the area term acquires a complex structure with both real and imaginary components. Since a complete analytic evaluation of the holographic tEE in higher-dimensional AdS–Schwarzschild backgrounds becomes increasingly intractable, we consider a more controlled regime in which the spacetime dimension is considered to be large. In this limit, we obtain a UV-finite and cutoff-independent expression for the timelike entanglement density, which separates into a real $\mathcal{O}(d^0)$ volume contribution and a complex $\mathcal{O}\left(\tfrac{1}{d}\right)$ area contribution.

Within the framework of holographic entanglement entropy, the area theorem \cite{Casini:2012ei,Casini:2016udt} yields a holographic realization of a \textit{weak} c-theorem along the renormalization group (RG) flow \cite{Casini:2006es,Casini:2012ei,Casini:2016udt,Liu:2012eea,Myers:2010tj,Myers:2010xs,Myers:2012ed}. This typically implies that the coefficient $\alpha$ of the area-law term takes a larger value in the ultraviolet than in the infrared, namely $\alpha_{\mathrm{UV}} \geq \alpha_{\mathrm{IR}}$. Under RG flow from the ultraviolet (UV) to the infrared (IR), the condition on $\alpha$ can be translated to the study of monotonic behavior of the entanglement density \cite{Gushterov:2017vnr}, involving the difference between the entanglement entropy in the excited and the ground states of the CFT. This behavior demonstrates that the finite part of the entanglement entropy provides a quantitative measure of the effective number of degrees of freedom in a quantum field theory, particularly along the renormalization group flow. However, in a strict sense, the area theorem has only been proven for spherical subsystems in spacetime dimension $d\geq 3$ \cite{Casini:2012ei,Casini:2016udt}. Interestingly, it has also been shown that for the spherical and the infinite stip subsystems, the area theorem fails in various spacetimes with specific dimensions and broken Lorentz symmetries \cite{Gushterov:2017vnr,Giataganas:2021jbj,Jain:2025xko,Chu:2019uoh,Baggioli:2020cld,Jokela:2025cyz}. In this context, timelike entanglement entropy emerges as a promising diagnostic as a holographic timelike $c$-function was recently proposed in \cite{Giataganas:2025div,Giataganas:2025ize} for generic spacetime dimensions, providing a probe of RG monotonicity in theories lacking Lorentz invariance. Following this work, it is natural to search for an analogous timelike area theorem involving the timelike entanglement entropy. To this end, we first define timelike entanglement density, an analogous quantity to the entanglement entropy density, to probe the timelike area theorem, where we look for a monotonic behavior from the UV to the IR. Interestingly, the whole study boils down to the computation of the coefficient of the area term. Since an exact computation becomes analytically prohibitive in generic spacetime dimension, we turn to the large-$d$ limit, where we find a clear violation of the monotonicity. This behavior mirrors the breakdown of the standard area theorem for spatial entanglement entropy, suggesting an intriguing parallel between timelike and spacelike entanglement structures in holography.

Taking a detour from the computation of timelike entanglement entropy (tEE), we investigate the geometric structure of the timelike and spacelike extremal surfaces in the near-horizon region of black holes. Related analyses in the context of holographic entanglement entropy have uncovered deep connections between the Ryu–Takayanagi (RT) surfaces and the chaotic properties of black holes, consistent with their role as the fastest scramblers of information \cite{Hayden:2007cs,Sekino:2008he,Susskind:2011ap,Lashkari:2011yi,Shenker:2013pqa,Maldacena:2015waa}. Classical chaos manifests through exponential sensitivity to initial conditions, whereas quantum chaos is commonly diagnosed by the growth of temporal operator commutators. In chaotic quantum systems, this growth typically follows $e^{\lambda_L (t - t_*)}$, where $\lambda_L$ is the Lyapunov exponent and $t_*$ the scrambling time. The fast scrambling dynamics of black holes then lead to the universal Maldacena–Shenker–Stanford bound $\lambda_L \leq \tfrac{2\pi}{\beta}$, with $\beta$ the inverse temperature \cite{Maldacena:2015waa,Jahnke:2019gxr,Malvimat:2021itk,Yuan:2020fvv}. When spatial separation is included, the commutator strength evolves as $e^{\lambda_L (t - t_* - |x|/v_B)}$, where $v_B$ is the butterfly velocity \cite{Hayden:2007cs,Sekino:2008he}. Holography provides a particularly effective framework for computing $v_B$ using entanglement wedge reconstruction where a boundary operator is represented by a bulk operator supported within the entanglement wedge, the region bounded by the RT surface and the boundary subsystem \cite{Shenker:2013pqa,Dong:2022ucb,Baishya:2024gih,Chua:2025vig}. Increasing the size of the boundary subsystem enlarges the corresponding entanglement wedge, effectively capturing the growth of the operator. Tracking this evolution offers a direct route to determining the butterfly velocity. Similarly, one may attempt to probe information growth along the temporal direction using timelike entanglement entropy. This, however, requires defining a timelike entanglement wedge, a construction that naturally confronts issues of causality. Recent progress has proposed such a definition, leading to related notions including the timelike entanglement wedge cross section \cite{Gong:2025pnu} and holographic timelike complexity \cite{Alishahiha:2025xml}. Building upon these developments, we analyze the growth of both spacelike and timelike extremal surfaces associated with the holographic tEE near the black hole horizon. Remarkably, we find that the exponential growths of both types of the extremal surfaces are governed by the same rate, saturating the MSS bound in the BTZ geometry as well as in black holes with Lifshitz anisotropy and hyperscaling violation. In the latter backgrounds, however, the effective Hawking temperature and consequently the Lyapunov exponent is modified by the nontrivial scaling exponents of the theory. Notably, unlike the standard entanglement entropy scenario, the timelike entanglement entropy exhibits this behavior for subsystems of arbitrary lengths, since the corresponding bulk surfaces always approach arbitrarily close to the horizon. These observations suggest that the bulk dual of tEE offers a meaningful probe of information growth in the temporal direction.

The plan of the paper is as follows. In \cref{sec_tee} we present a generic construction of the holographic timelike entanglement entropy, highlighting the emergence of the real and the imaginary components from spacelike and timelike extremal surfaces, respectively. In \cref{sec_tee_btz} we apply this construction to the BTZ black hole background, where both the timelike and spacelike surfaces can be solved analytically. In \cref{sec_tee_d_bh} we generalize the analysis to higher-dimensional AdS–Schwarzschild black holes, identifying a critical turning point for the timelike surface, and examine the volume–area structure of the holographic tEE as well as the timelike area theorem at large subsystem length and at large spacetime dimensions. In \cref{sec_tee_growth} we investigate the near-horizon behavior of the extremal surfaces constituting the holographic tEE in different black-hole geometries. We determine their exponential growth rates in various black hole backgrounds. Finally, \cref{dis} concludes with a summary of our findings and discusses the physical implications of the holographic tEE for temporal correlations and information growth in black-hole backgrounds.

\section{Holographic timelike entanglement entropy}\label{sec_tee}
In this section, we briefly review the holographic construction of the timelike entanglement entropy (tEE) \cite{Doi:2022iyj,Doi:2023zaf,Basak:2023otu,Afrasiar:2024lsi,Afrasiar:2024ldn}. The holographic tEE can be obtained through a pair of spacelike and a timelike extremal surfaces. The areas of the spacelike and the timelike surfaces yield real and imaginary values, respectively, which correspond to the holographic tEE. The pair of spacelike surfaces extend from the endpoint of the boundary temporal subsystem to deep into the bulk gravity region, whereas the timelike surface stretches from a specific turning point until the deep interior of the bulk. The choice of the turning point is specifically dependent on the size of the temporal subsystem in question. The domains of validity of these extremal surfaces in the bulk gravity region is determined by the reality conditions of their corresponding equations of motion.

To elaborate mathematically, we consider a generic gravity background, 
\begin{align}
    ds^2_{d+1} = g_{tt}(r)dt^2+g_{rr}(r)dr^2+g_{xx}(r)dx_{d-1}^2~,
\end{align}
where $t$ is the Lorentzian time direction and $r$ is the holographic direction of the gravity region. In the above metric, the spacetime possesses $(d-1)$ spatial directions, which are denoted by the $x$ coordinates. The asymptotic boundary of this gravity background is considered at $r \rightarrow \infty$ where a timelike strip subsystem $A \equiv -T/2 \leq t \leq T/2$ is considered on a constant spatial $x$-slice and extends infinitely along the transverse $x_{d-2}$ directions. However, for the sake of the computation, we will consider the lengths of the strip subsystem in all the transverse directions to be $L$s, where in principle, it is infinitely long. The holographic tEE can be computed for this subsystem $A$ by extremizing the area integral as,
\begin{align}\label{area_gen}
    4 G_N^{(d+1)} \mathcal{S}^T=2 L^{d-2} \int_{r_d}^{r_u}dr~ g_{xx}^{\frac{d-2}{2}}\sqrt{g_{rr}+g_{tt} t^{\prime}(r)^2}~.
\end{align}
Utilizing this extremal area expression, one can obtain the Euler-Lagrange equation of motion for the extremal surfaces extended into the bulk region as
\begin{align}\label{EOM_gen}
    t^{\prime}(r)^2 = \frac{c_0^2\, g_{rr}}{g_{tt} \left(g_{tt} \, g_{xx}^{d-2} - c_0^2\right)}~,
\end{align}
with $c_0$ being the constant of integration. The length of the temporal subsystem $T$ at the asymptotic boundary can be obtained by integrating \cref{EOM_gen} with respect to the holographic direction
\begin{align}\label{Tint}
    T= 2 \int_{r_d}^{r_u} dr~ t^{\prime}(r)~.
\end{align}
In the above expression, $r_{d}$ and $r_{u}$ denote the bulk radial coordinates at which the corresponding extremal surfaces terminate, as determined by the equation of motion in \cref{EOM_gen}. The temporal subsystem lengths will be evaluated in terms of bulk geometric data using \cref{Tint}, for all admissible surfaces allowed by the equations of motion. Since this computation of the subsystem length depends on the background geometry, we will specify the relevant details in a later section. 

From a mathematical point of view, one can now have the liberty to choose the sign of the constant $c_0^{2}$ in \cref{EOM_gen}, allowing it to be either positive or negative, and consequently, we define $c_0^2=s\,C^2$ with $s=\pm 1$ and $C^2>0$. For the case $s = -1$, the corresponding extremal surface possesses a turning point at $r = r_{0}$ in the bulk gravity region, defined by $c_0^2=-C^2=g_{tt}(r) g_{xx}(r)^{d-2}|_{r=r_0}$. We describe the corresponding equation of motion by $t^{\prime}_{\text{Im}}(r)$ and subsequently, upon imposition of the reality condition on $t^{\prime}_{\text{Im}}(r)$, one can argue that this surface can only exist inside the bulk region and never reaches the asymptotic boundary. By denoting the corresponding hypersurface as $\Sigma_{\text{Im}}$, we examine the norm of its normal vector, which satisfies the condition $g^{\mu\nu}\partial_\mu \Sigma_{\text{Im}}\partial_\nu \Sigma_{\text{Im}} > 0$, implying that $\Sigma_{\text{Im}}$ is a timelike hypersurface. Furthermore, in the subsequent sections, we will demonstrate that this hypersurface encodes the imaginary component of the holographic tEE, thereby justifying the notation ``Im" in the suffix. The corresponding extremal area integral for this surface can be expressed as
\begin{align}\label{area_gen_Im}
    4 G_N^{(d+1)} \mathcal{S}_{\text{Im}}^T=2 L^{d-2} \int_{r_d}^{r_0}dr~ g_{xx}^{\frac{d-2}{2}}\sqrt{g_{rr}+g_{tt} t^{\prime}_{\text{Im}}(r)^2}~.
\end{align}
There can also be another extremal surface for $s=+1$ with $c_0^2=C^2=-g_{tt}(r) g_{xx}(r)^{d-2}|_{r=r_0}$ for which we cannot associate a turning point following \cref{EOM_gen}. The corresponding equation of motion is now denoted by $t^{\prime}_{\text{Re}}(r)$ and depending upon the reality condition, it indicates two pieces of extremal surfaces with a mirror symmetry extended from the boundary of the temporal subsystem to the deep bulk region. As earlier, we denote the associated hypersurfaces by $\Sigma_{\text{Re}}$ and evaluate the norm of their normal vectors, which satisfies the condition $g^{\mu\nu}\partial_\mu \Sigma_{\text{Re}}\partial_\nu \Sigma_{\text{Re}} < 0$, indicating that $\Sigma_{\text{Re}}$ corresponds to the spacelike hypersurfaces. Later in the following sections, we will show that the total area of these spacelike hypersurfaces contributes to the real part of the holographic tEE, thereby justifying the suffix ``Re". We can now express the corresponding extremal area integral for these surfaces as
\begin{align}\label{area_gen_Re}
    4 G_N^{(d+1)} \mathcal{S}_{\text{Re}}^T=2 L^{d-2} \int_{r_d}^{\infty}dr~ g_{xx}^{\frac{d-2}{2}}\sqrt{g_{rr}+g_{tt} t^{\prime}_{\text{Re}}(r)^2}~.
\end{align}
These spacelike surfaces contain UV divergences from the asymptotic boundary at $r=\infty$, which can be renormalized from a pair of straight disconnected surfaces satisfying the equation of motion $t^{\prime}(r)=0$. The corresponding area integral for these surfaces can be written as
\begin{align}\label{area_discgen_Re}
    4 G_N^{(d+1)} \mathcal{S}_{\text{discon}}^T=2 L^{d-2} \int_{0}^{\infty}dr~ g_{xx}^{\frac{d-2}{2}}\sqrt{g_{rr}}~.
\end{align}
This disconnected solution is extended from the asymptotic boundary $r\to\infty$ all the way up to the singularity $r=0$ of the bulk geometry. However, in our present article, we will not explicitly use the renormalization of the holographic tEE with \cref{area_discgen_Re} in our analytic computations. Nevertheless, this renormalization procedure will be employed in the numerical evaluation of the extremal surface areas, as it provides better control and stability in the analysis.

\section{BTZ black hole}\label{sec_tee_btz}
Previously, in \cite{Afrasiar:2024ldn,Afrasiar:2024lsi}, the holographic construction of the timelike entanglement entropy was investigated either in standard anisotropic spacetimes \cite{Afrasiar:2024ldn} or in spacetime with a natural geometric endpoint characterized by a cigar-like bulk geometry \cite{Afrasiar:2024lsi}. In this work, we extend the framework to spacetimes containing a horizon, which naturally partitions the bulk into distinct regions. A canonical example of such a background is the BTZ black hole in $(2+1)$-dimensional AdS gravity. Within this geometry, we explicitly construct both the spacelike and timelike extremal surfaces governed by the equations of motion introduced in the previous section and analyze their behavior across the bulk regions. The AdS BTZ black hole metric in Lorentzian signature is described by
\begin{align}\label{btz_geo}
    ds^2=-\left(r^2-r_h^2\right)dt^2+\frac{dr^2}{r^2-r_h^2}+r^2 d\phi^2~,
\end{align}
where $r_h$ is the radius of the horizon. The asymptotic boundary of the metric in \cref{btz_geo} is located at $r\rightarrow \infty$ where the dual CFT$_2$ at a finite temperature resides. The temporal subsystem $A\equiv-T/2\leq t\leq T/2$ is considered in the Lorentzian time direction $t$ and at a constant $\phi$ slice. Here the the timelike entanglement entropy will be computed by minimizing the area integral,
\begin{align}
    4 G_N^{(d+1)} \mathcal{S}^T=2 L^{d-2} \int_{r_d}^{r_u}dr~ \sqrt{\left(r^2-r_h^2\right)^{-1}-\left(r^2-r_h^2\right) t^{\prime}(r)^2}~.
\end{align}
The equations of motion for the timelike and spacelike surfaces can be read from \cref{EOM_gen} as
\begin{align}
    {t^{\prime}_{\text{Im}}}(r)^2 &= \frac{r_h^2-r_0^2}{\left(r^2-r_0^2\right) \left(r^2-r_h^2\right)^2}~,\label{t_prime_im}\\
    {t^{\prime}_{\text{Re}}}(r)^2 &=\frac{r_0^2-r_h^2}{\left(r^2+r_0^2-2 r_h^2\right) \left(r^2-r_h^2\right)^2}~,\label{t_prime_re}
\end{align}
where the constants of the integrations are considered as $c_0=\sqrt{r_h^2-r_0^2}$ and $c_0=\sqrt{-r_h^2+r_0^2}$ for the timelike and the spacelike surfaces respectively. Note that, in the above set of equations, $r_0$ is the turning point corresponding to the timelike surface which is uniquely fixed for a particular length of the temporal subsystem at the boundary.  
The solutions described in \cref{t_prime_im,t_prime_re} exist in both the regions outside and inside the black hole horizon. The validity of the spacelike and timelike surfaces can be determined by examining the regions where their respective equations of motion remain real. Such a validity check indicates two ranges of $r_0$ as $r_0\geq r_c$ and $r_0<r_c$ for $r_c=\sqrt{2}r_h$ which yield two completely different equations of motion. Furthermore the region $r_0<r_c$ can further be broken into $0<r_0<r_h$ and $r_h<r_0<r_c$ as their areas demonstrate different behaviours, although the equations of motion show similar characteristics. Below we will consider these regions separately and explore the behaviour of the timelike entanglement entropy.

\subsection{Bulk region I: $\bm{r_0\geq r_c}$}\label{regI}
\begin{figure}[h]
	\centering
	\includegraphics[width=.55\linewidth]{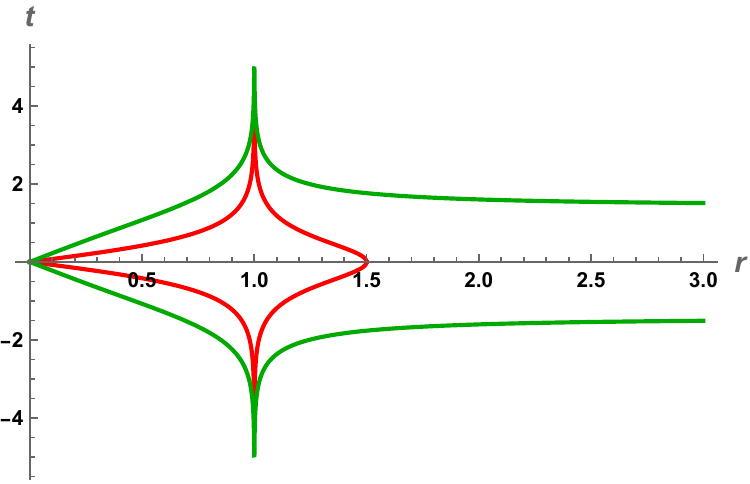}
	\caption{Holographic timelike entanglement entropy for BTZ black hole background with $r_h=1$ and the turning point $r_0=1.5$. The spacelike (green) and the timelike surfaces are demonstrated. }  
	\label{surfteebh}
\end{figure}

In this region, we find that the equation of motion for the timelike surface remains valid from the turning point $r_{0}$, located outside the horizon at $r_{h}$, until the singularity at $r=0$. In contrast, the spacelike surfaces extend from the asymptotic boundary $r \rightarrow \infty$ all the way up to the deep infrared (IR) region $r=0$ as depicted in \cref{surfteebh}. 
The surfaces in \cref{t_prime_im,t_prime_re} satisfy the condition $t'_{\text{Re}}(r)\big{|}_{r\rightarrow r_h^{\pm}}=t'_{\text{Im}}(r)\big{|}_{r\rightarrow r_h^{\pm}}=\pm \infty$ which indicate their continuity at the horizon. Consequently, the analytic evaluation of the subsystem length $T$ at the asymptotic boundary can be naturally separated into contributions from the regions outside and inside the black hole horizon $r_h$, which are then combined to yield the final expressions. Considering first the black hole interior region, we obtain the subsystem lengths by integrating \cref{t_prime_im,t_prime_re} with respect to the bulk direction as
\begin{align}
    T_{\text{Im,inside}}&= 2 \int_{0}^{r_h(1-\epsilon_1)} dr~ t^{\prime}_{\text{Im}}(r) =\frac{1}{r_h}\log \left(\frac{r_0^2 \epsilon_1}{2 \left(r_0^2-r_h^2\right)}\right)~,\notag\\
    T_{\text{Re,inside}}&= 2 \int_{0}^{r_h(1-\epsilon_1)} dr~ t^{\prime}_{\text{Re}}(r) = \frac{1}{r_h}\log \left(\frac{\left(r_0^2-2 r_h^2\right)\epsilon_1}{ 2 \left(r_0^2-r_h^2\right)}\right)~,
\end{align}
where, $\epsilon_1 \ll 1$ is introduced to keep track of the infinities arising from the horizon $r_h$.
Therefore, the contribution to the subsystem length from the extremal surfaces inside the horizon,
\begin{align}\label{T_in}
   T_{\text{in}}= T_{\text{Im,inside}}-T_{\text{Re,inside}}=\frac{1}{r_h}\log \left(\frac{r_0^2}{r_0^2-2 r_h^2}\right)~,
\end{align}
where the $\epsilon_1$ divergent contributions are canceled out.
Unlike the \cref{surfteebh}, \cref{T_in} clearly demonstrate a difference $T_{\text{in}}$ between the spacelike and the timelike surfaces. Note that, the plots in \cref{surfteebh} are obtained by solving the first order differential equations of $t^\prime_{\text{Im}}(r)$ and $t^\prime_{\text{Re}}(r)$ with imposing the homology condition. Similar to the above analysis, we obtain the subsystem lengths from the extremal surfaces extending at the exterior of the black hole by integrating \cref{t_prime_im,t_prime_re} with respect to the bulk direction,
\begin{align}
    T_{\text{Im,outside}}&=2 \int_{r_h(1+\epsilon_2)}^{r_0} dr~ t^{\prime}_{\text{Im}}(r) = \frac{1}{r_h}\log \left(\frac{2 \left(r_0^2-r_h^2\right)}{r_0^2 \epsilon_2}\right)~,\notag\\
    T_{\text{Re,outside}}&= 2\int_{r_h(1+\epsilon_2)}^{\infty} dr~ t^{\prime}_{\text{Re}}(r) = \frac{1}{r_h}\log \left(\frac{2 \left(r_0^2-r_h^2\right) \left(r_0^2-2 r_h \sqrt{r_0^2-r_h^2}\right)}{\epsilon_2\left(r_0^2-2 r_h^2\right){}^2}\right)~,
\end{align}
where as earlier $\epsilon_2 \ll 1$ is introduced to keep track of the infinities arising from the horizon $r_h$.
Therefore, the contribution to the subsystem length from the extremal surfaces existing outside the horizon,
\begin{align}\label{T_out}
   T_{\text{out}}= T_{\text{Im,outside}}-T_{\text{Re,outside}}=\frac{1}{r_h}\log \left(1+ \frac{2r_h \sqrt{r_0^2-r_h^2}}{r_0^2}\right)~.
\end{align}
Similar to \cref{T_in}, the $\epsilon_2$ divergent contributions are canceled out in the above expression.
We now define the total subsystem length to be
\begin{align}
T &=T_{\text{out}} + T_{\text{in}}\label{T_tot1}\\
   &=\frac{1}{r_h}\log \left(\frac{r_0^2 + 2 r_h \sqrt{r_0^2-r_h^2}}{r_0^2-2 r_h^2}\right)~\label{T_tot}.
\end{align}
Note that in the BTZ black hole background, we do not observe a patching between the spacelike and timelike surfaces according to the patching condition described in \cite{Afrasiar:2024lsi}. The expression \cref{T_tot} shows that the total subsystem length $T$ increases as the turning point $r_0$ is pushed deeper into the bulk, eventually diverging at a critical value $r_c=\sqrt{2}r_h$.
Furthermore, one can also note that if the turning point $r_0$ is shifted towards the asymptotic boundary $r\rightarrow\infty$, the subsystem length decreases and finally vanishes. Inverting \cref{T_tot}, the turning point $r_0$ can be expresses in terms of the total subsystem length $T$ as
\begin{align}\label{r0_T}
   r_0^2 = r_h^2 \frac{\cosh \left(r_h T\right) }{\sinh^2{\left(\frac{r_h T}{2}\right)}}~.
\end{align}
This expression of the turning point will be utilized in the computation for the area of the spacelike and the timelike surfaces in terms of the subsystem length.

Utilizing \cref{area_gen_Im} and \cref{area_gen_Re}, we can obtain the areas of the timelike and the spacelike surfaces that exist both inside and outside the horizon as
\begin{align}\label{tEE_r0}
    4G_N^{(3)}\mathcal{S}_{\text{Im}}^T &= 2\int_{0}^{r_0} dr~ \frac{1}{\sqrt{r^2-r_0^2}} = i \pi ~,\notag\\
    4G_N^{(3)}\mathcal{S}_{\text{Re}}^T &= 2\int_{0}^{1/\epsilon} dr~ \frac{1}{\sqrt{r^2+r_0^2-2 r_h^2}} = 2 \log \left(\frac{2}{\epsilon  \sqrt{r_0^2-2 r_h^2}}\right)~,
\end{align}
where $\epsilon$ is the UV cut-off introduced at the asymptotic boundary $r=\infty$. The total timelike entanglement entropy (tEE) can now be expressed as a sum of the spacelike and timelike extremal areas,
\begin{align}\label{tEE_r0_tot}
   \mathcal{S}^T(A) = \mathcal{S}_{\text{Re}}^T + \mathcal{S}_{\text{Im}}^T &= \frac{2}{4G_N^{(3)}}  \log \left(\frac{2}{\epsilon  \sqrt{r_0^2-2 r_h^2}}\right) + \frac{i \pi}{4G_N^{(3)}}~,\notag\\
   &= \frac{c}{3}  \log \left(\frac{2}{\epsilon  \sqrt{r_0^2-2 r_h^2}}\right) + i \frac{\pi c}{6}~,
\end{align}
where in the last equality we have utilized the Brown-Henneaux formula $c=\frac{3}{2G_N^{(3)}}$ with $c$ being the central charge of the dual CFT$_2$ at the asymptotic boundary \cite{Brown:1986nw}. Subsequently, we now utilize \cref{r0_T} to express the tEE in terms of the total subsystem length $T$ as
\begin{align}\label{tEE_T}
   \mathcal{S}^T(A) = \frac{c}{3} \log \left[\frac{2}{\epsilon \,  r_h} \sinh \left(\frac{r_h T}{2}\right)\right] + i \frac{\pi c}{6}~,
\end{align}
which matches exactly with the earlier results in the literature \cite{Doi:2023zaf}.

We will conclude this subsection with an investigation of the timelike entanglement entropy for a small subsystem ($r_0\to \infty$) at the boundary. Simultaneously, a similar scenario can be considered for a small black hole temperature or horizon radius. Both of these situations can be understood as $\frac{r_h}{r_0}$ or $r_hT$ approaching zero. Applying this limit in \cref{t_prime_im} and \cref{t_prime_re}, it can be observed that the equation of motions involves the presence of the horizon perturbatively as
\begin{align}
    {t^{\prime}_{\text{Im}}}(r)^2 &= \frac{r_0^2}{r^4 \left(r_0^2-r^2\right)}+\frac{r_0^2 \left(r^2-2 r_0^2\right) }{r^6\left(r^2- r_0^2\right)}\left(\frac{r_h}{r_0}\right)^2+\mathcal{O}\left(\frac{r_h}{r_0}\right)^4,\notag\\
    {t^{\prime}_{\text{Re}}}(r)^2 &= \frac{r_0^2}{r^4 \left(r^2+r_0^2\right)}+\frac{r_0^2 \left(-r^4+3 r_0^2 r^2+2 r_0^4\right)}{r^6 \left(r^2+r_0^2\right)^2}\left(\frac{r_h}{r_0}\right)^2+\mathcal{O}\left(\frac{r_h}{r_0}\right)^4~,
\end{align}
where the first terms yield the same result as the vacuum AdS$_3$ \cite{Basak:2023otu,Afrasiar:2024lsi}. Furthermore a similar characteristics can be obtained for the timelike entanglement entropy as
\begin{align}
   \mathcal{S}^T(A) =\frac{c}{3}  \log \left(\frac{T}{\epsilon }\right) +\frac{c}{12} \left(r_hT\right)^2 + i \frac{\pi c}{6} +\mathcal{O}\left(r_hT\right)^4~.
\end{align}
In the opposite limit where $r_hT\to\infty$ the timelike entanglement entropy can be expressed as,
\begin{align}
   \mathcal{S}^T(A) =\frac{c}{3}  \log \left(\frac{1}{\epsilon r_h}\right) +\frac{c~r_h T}{6} + i \frac{\pi c}{6} +\mathcal{O}\left(e^{-r_hT}\right)~.
\end{align}

\subsection{Bulk region II: $\bm{r_h\leq r_0 \leq r_c}$}\label{regII}
\begin{figure}[h]
	\centering
	\includegraphics[width=.55\linewidth]{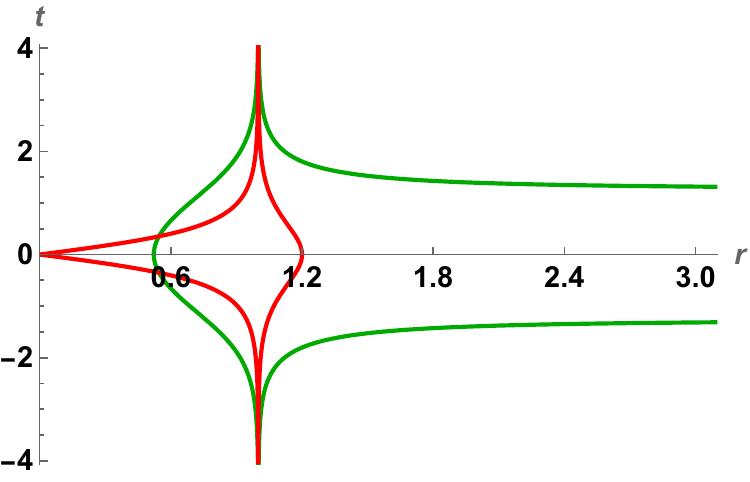}
	\caption{Depiction of the extremal surfaces associated with the holographic timelike entanglement entropy (tEE) in the BTZ black hole geometry, corresponding to a different position of the turning point $r_0$ relative to the black hole horizon $r_h$. This configuration is characterized by the parameter regime $r_h < r_0 < r_c$, with $r_h = 1$ and $r_0 = 1.2$.}  
	\label{r0_lt_rc_gt_rh}
\end{figure}

In this regime, the turning point $r_0$ lies between the black hole horizon $r_h$ and the critical turning point $r_c=\sqrt{2}r_h$. Within this parameter range, the corresponding timelike and spacelike extremal surfaces are determined by the equations of motion given in \cref{t_prime_im,t_prime_re}. Note that for $r_h < r_0 < r_c$, the spacelike surface can now develop a turning point in the bulk gravitational region at $r = \sqrt{2r_h^2-r_0^2}$. In particular, the spacelike surface extends from this turning point at $r = \sqrt{2r_h^2-r_0^2}$ to the asymptotic boundary at $r \to \infty$, while the timelike surface continues to reside entirely within the interior region, bounded between the singularity at $r = 0$ and the turning point $r_0$. The extremal surfaces corresponding to this region are plotted in \cref{r0_lt_rc_gt_rh}. As shown in \cref{r0_lt_rc_gt_rh}, both the surfaces exhibit turning points at the interior and the exterior of the horizon, corresponding to the timelike and the spacelike surfaces, respectively. Although these surfaces cross each other in the bulk and thus fail to satisfy as the bulk dual of the timelike entanglement entropy, we will show some of the interesting characteristics of these surfaces. Similar to \cref{regI}, we compute $T_{\text{in}}$ and $T_{\text{out}}$ as 
\begin{align}
    T_{\text{in}}= T_{\text{Im,inside}}-T_{\text{Re,inside}}
    &=2 \int_{0}^{r_h(1-\epsilon_1)} dr~ t^{\prime}_{\text{Im}}(r)-2 \int_{\sqrt{2 r_h^2-r_0^2}}^{r_h(1-\epsilon_1)} dr~ t^{\prime}_{\text{Re}}(r)\notag\\
    &=\frac{1}{r_h}\log \left(-\frac{r_0^2}{r_0^2-2 r_h^2}\right)~,
\end{align}
\begin{align}
    T_{\text{out}}= T_{\text{Im,outside}}-T_{\text{Re,outside}}
    &=2 \int_{r_h(1+\epsilon_2)}^{r_0} dr~ t^{\prime}_{\text{Im}}(r)-2 \int_{r_h(1+\epsilon_2)}^{\infty} dr~ t^{\prime}_{\text{Re}}(r)\notag\\
    &=\frac{1}{r_h}\log \left(1+\frac{2 r_h\sqrt{r_0^2-r_h^2}}{r_0^2}\right)~,
\end{align}
which can be combined to compute the subsystem length as a function of the turning point of the timelike surfaces as, 
\begin{align}\label{TII}
    T=T_{\text{in}}+T_{\text{out}}=\frac{1}{r_h}\log \left(-\frac{r_0^2+2r_h \sqrt{r_0^2-r_h^2}}{r_0^2-2 r_h^2}\right)~.
\end{align}
Note that, the subsystem size decreases gradually to zero when the turning point $r_0$ approaches the horizon $r_h$. On the other hand, the subsystem length diverges as $r_0$ moves towards the critical point $r_c=\sqrt{2}r_h$. As earlier, the turning point $r_0$ from the above expression can be expressed in terms of the subsystem length $T$ as
\begin{align}\label{r0_TII}
   r_0^2 = r_h^2 \left( 1+ \tanh^2\left(\frac{r_h T}{2} \right)\right)~.
\end{align}
The sum of the areas of the spacelike and the timelike surfaces in terms of the subsystem length now yields
\begin{align}\label{STII}
   \mathcal{S}_{\text{II}}^T(A) &= \frac{c}{3}  \log \left(-\frac{2}{\epsilon  \sqrt{r_0^2-2 r_h^2}}\right) + i \frac{\pi c}{6}=\frac{c}{3} \log \left[\frac{2}{\epsilon \,  r_h} \cosh \left(\frac{r_h T}{2}\right)\right] + i \frac{\pi c}{6}~,
\end{align}
where, in the last equality, we have utilized \cref{r0_TII}. However, we will discard the solutions in this subsection as the homology condition is violated. It is important to emphasize that the final expression of the total area obtained in this configuration does not reproduce the tEE computed from the dual CFT side.

\subsection{Bulk region III: $\bm{0 \leq r_0 \leq r_h}$}
In this regime, the turning point $r_0$ lies behind the horizon $r_h$, leading to an inversion of the characteristic behavior of the extremal surfaces governed by the equations of motion given in \cref{t_prime_im,t_prime_re}. Specifically, by examining the norm of the normal vector, $g^{\mu\nu}\partial_\mu \Sigma \partial_\nu \Sigma$, for the hypersurfaces $\Sigma_{\text{Im}}$ and $\Sigma_{\text{Re}}$, we find that the characteristics of the extremal surfaces associated with \cref{t_prime_im,t_prime_re} become interchanged. Consequently, in this region, the spacelike and the timelike surfaces are described by
\begin{align}
    {t^{\prime}_{\text{Re}}}^{\text{III}}(r)^2 &= \frac{r_h^2-r_0^2}{\left(r^2-r_0^2\right) \left(r^2-r_h^2\right)^2}~,\label{t_prime_re_3}\\
    {t^{\prime}_{\text{Im}}}^{\text{III}}(r)^2 &=\frac{r_0^2-r_h^2}{\left(r^2+r_0^2-2 r_h^2\right) \left(r^2-r_h^2\right)^2}~\label{t_prime_im_3}.
\end{align}
Interestingly, from the above expressions, we note that in this region, both the spacelike and timelike surfaces exhibit individual turning points in the bulk. In particular, the spacelike surface possesses a turning point at $r = r_0$, located inside the horizon, whereas the timelike surface exhibits a turning point at $r = \sqrt{2r_h^2 - r_0^2}$ outside the horizon. Although the schematic illustration in \cref{r0_lt_rc_gt_rh} depicts the qualitative behavior of the extremal surfaces corresponding to \cref{regII}, the extremal surfaces in this regime exhibit an identical behavior, with appropriately redefined turning points associated with the spacelike and the timelike surfaces described above. Notably, however, the two extremal surfaces again intersect within the black hole interior, which violates the required homology condition for a consistent holographic construction of tEE. Consequently, we do not regard this configuration as a valid candidate for the holographic time-entanglement surface. However, we will present a computation parallel to \cref{regI,regII} where $T_{\text{in}}$ and $T_{\text{out}}$ can be expressed as earlier
\begin{align}
    T_{\text{in}}= T_{\text{Im,inside}}-T_{\text{Re,inside}}
    &=2 \int_{0}^{r_h(1-\epsilon_1)} dr~ {t^{\prime}_{\text{Im}}}^{\text{III}}(r)-2 \int_{r_0}^{r_h(1-\epsilon_1)} dr~ {t^{\prime}_{\text{Re}}}^{\text{III}}(r)\notag\\
    &=\frac{1}{r_h}\log \left(-1+\frac{2 r_h^2}{r_0^2}\right)~.
\end{align}
\begin{align}
    T_{\text{out}}= T_{\text{Im,outside}}-T_{\text{Re,outside}}
    &=2 \int_{r_h(1+\epsilon_2)}^{\sqrt{2 r_h^2-r_0^2}} dr~ {t^{\prime}_{\text{Im}}}^{\text{III}}(r)-2 \int_{r_h(1+\epsilon_2)}^{\infty} dr~ {t^{\prime}_{\text{Re}}}^{\text{III}}(r)\notag\\
    &=\frac{1}{r_h}\log \left(1-\frac{2 r_h \sqrt{r_h^2-r_0^2}}{r_0^2-2 r_h^2}\right)~.
\end{align}
Finally, the length of the subsystem reads as,
\begin{align}\label{TIII}
    T=T_{\text{in}}+T_{\text{out}}=\frac{1}{r_h}\log \left(-1+\frac{2 r_h \left(r_h+\sqrt{r_h^2-r_0^2}\right)}{r_0^2}\right)~.
\end{align}
Note that, the subsystem length vanishes and becomes infinity as the limits $r_0\to r_h$ and $r_0\to 0$, respectively. Inverting the above equation, we express the turning point in terms of the subsystem length
\begin{align}\label{r0III}
    r_0^2=\frac{r_h^2}{\cosh^2{\left(\frac{r_hT}{2}\right)}}~.
\end{align}
Furthermore, we compute the sum of the areas of the spacelike and the timelike surfaces as,
\begin{align}\label{tEE_II}
   \mathcal{S}_{\text{III}}^T(A)&= \frac{c}{3}\log \left(\frac{2}{\epsilon \, r_0}\right)+i \frac{\pi c}{6}=\frac{c}{3} \log \left[\frac{2}{\epsilon \,  r_h} \cosh \left(\frac{r_h T}{2}\right)\right] + i \frac{\pi c}{6}~.
\end{align}
where in the second equality, we substitute the $r_0$ using \cref{r0III}. Note that, similar to the \cref{regII}, the total area in \cref{tEE_II} does not reproduce the result obtained from the dual CFT side.

\section{AdS-Schwarzschild black hole}\label{sec_tee_d_bh}
Here, we consider a $(d+1)$-dimensional AdS-Schwarzschild black hole background with the metric
\begin{align}\label{sschild_high_d}
   ds_{d+1}^2=-r^2 f(r)dt^2+\frac{dr^2}{r^2 f(r)}+r^2\left(dy^2+dx_{d-2}^2\right).
\end{align}
where $f(r)=1-\left(\frac{r_h}{r}\right)^d$ is the blackening factor. The asymptotic boundary of this geometry is situated at $r \to \infty$, where the dual CFT$_d$ is defined. In this background, we consider a strip-like subsystem $A \equiv -T/2 \leq t \leq T/2$ in the boundary theory, specified along the Lorentzian time direction $t$, at a fixed spatial coordinate $y$, while being infinitely extended along the remaining transverse spatial directions $x_{d-2}$. The extremized area integral can be read from \cref{area_gen} as
\begin{align}\label{area_int}
   4 G_N^{(d+1)} \mathcal{S} = L^{d-2} \int_{r_d}^{r_u} dr~ r^{d-2} \sqrt{r^{-2} \left(1-\left(\frac{r_h}{r}\right)^d\right)^{-1}-r^2 \left(1-\left(\frac{r_h}{r}\right)^d\right) t'(r)^2}~.
\end{align}
Here $L^{d-2}$ denotes the $(d-2)$-dimensional volume of the remaining transverse spatial $x$-directions, and $G_N^{(d+1)}$ is the Newton’s constant of the $(d+1)$-dimensional AdS geometry. Following \cref{sec_tee}, we extremize \cref{area_int} and obtain the equations of motion of the bulk surfaces constituting the holographic tEE as,
\begin{align}\label{t_prime_wbh}
    {t^{\prime}_{\text{Im}}}(r)^2 &= \frac{r_0^{2 d} \left(1-\left(\frac{r_h}{r_0}\right)^d\right)}{r^2 \left(\left(\frac{r_h}{r}\right)^d-1\right)^2 \left(r_0^2 r^d r_h^d+r^2 r_0^d \left(r_0^d-r_h^d\right)-r_0^2 r^{2 d}\right)}~,\notag\\
    {t^{\prime}_{\text{Re}}}(r)^2 &=\frac{r_0^{2 d} \left(1-\left(\frac{r_h}{r_0}\right)^d\right)}{r^2 \left(1-\left(\frac{r_h}{r}\right)^d\right)^2 \left(-r_0^2 r^d r_h^d+r^2 r_0^d \left(r_0^d-r_h^d\right)+r_0^2 r^{2 d}\right)}~,
\end{align}
where $r_0$ is the turning point of the timelike surface. Similar to the BTZ black hole scenario, we can broadly identify three different solution regions depending on the positions of the turning point $r_0$.

\begin{figure}[]
	\centering
	\includegraphics[width=.55\linewidth]{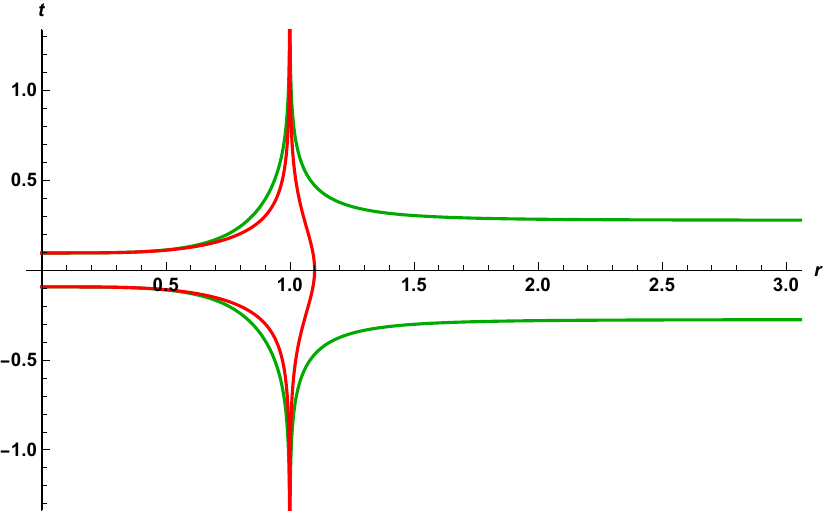}
	\caption{The spacelike (green) and the timelike (red) surfaces are illustrated in $AdS_6$-Schwarzschild black hole background. The black hole horizon is $r_h=1$ and the turning point of the timelike surface is $r_0=1.1$ which situates in the region $r_0>r_c$. The spacelike surface is extended from the boundary to the interior of the black hole. The timelike and the spacelike surfaces are merged at $r=0$ where the merging condition $g^{\m\n}\pp_\m \S_{\text{Re}}\pp_\n \S_{\text{Re}}|_{r=0}=g^{\m\n}\pp_\m \S_{\text{Im}}\pp_\n \S_{\text{Im}}|_{r=0}=0$ is satisfied. 
    }  
	\label{surfteewbh}
\end{figure}

We observe from \cref{t_prime_wbh} that, for sufficiently large values of the turning point $r_0 (>r_h)$, the condition $t^{\prime}_{\text{Re}}(r)=\infty$ admits no real solution other than the trivial one at the horizon, $r=r_h$. As a consequence, the spacelike extremal surfaces do not contain any turning point in the bulk. Instead, these surfaces extend smoothly from the asymptotic boundary at $r\to\infty$ all the way up to the curvature singularity at $r=0$ as depicted in \cref{surfteewbh}. However, the equations of motion in this region show that the timelike surface extends from the turning point $r_0$ to the black hole singularity at $r=0$. Investigating the norm of the normal vectors $g^{\mu\nu}\pp_\m \S\pp_\n \S$ for both the extremal surfaces, where $\S$ indicates the hypersurfaces corresponding to the spacelike and the timelike surfaces, we observe that both the surfaces become lightlike at $r=0$. Thus we merge them at $r=0$ utilizing the merging condition $g^{\m\n}\pp_\m \S_{\text{Re}}\pp_\n \S_{\text{Re}}|_{r=0}=g^{\m\n}\pp_\m \S_{\text{Im}}\pp_\n \S_{\text{Im}}|_{r=0}=0$ \cite{Afrasiar:2024lsi}. This property shows that the bulk dual of the timelike entanglement entropy consists of the timelike and the spacelike surfaces which are merged smoothly inside the bulk as shown in \cref{surfteewbh}.

\begin{figure}[h]
	\centering
	\includegraphics[width=.55\linewidth]{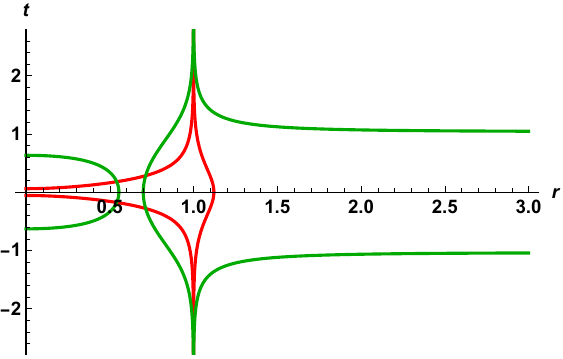}
	\caption{We demonstrate the spacelike (green) and timelike (red) surfaces in $AdS_4$-Schwarzschild spacetime with the black hole horizon $r_h=1$ and the turning point of the timelike surface $r_0=1.12$. The turning point satisfies the condition $r_0<r_c$. Here we observe, crossing spacelike and timelike surfaces which makes the solution invalid.}  
	\label{surfteewbhreg2}
\end{figure}

As the turning point $r_0$ is shifted deeper into the bulk and approaches the horizon $r_h$, we find a qualitative change in the structure of solutions. In particular, there exists a critical value $r_0=r_c$ such that, for $r_0<r_c$, additional real solutions to the condition $t^{\prime}_{\text{Re}}(r)=\infty$ emerge along with the trivial one at $r=r_h$. This indicates that, in the parameter range $r_h<r_0<r_c$, the spacelike extremal surfaces necessarily develop turning points within the bulk gravitational region. A pictorial depiction of the extremal surfaces in this regime is described in \cref{surfteewbhreg2}.

Subsequently, when $r_0$ is pushed further to be located inside the horizon, $r_0<r_h$, the extremal surfaces exhibit an additional qualitative transition in their geometric behavior. In this regime, the equations of motion governing the spacelike and timelike extremal surfaces, as given in \cref{t_prime_wbh}, are effectively interchanged. As a consequence, the turning point $r_0$ is now naturally associated with spacelike extremal surfaces, rather than with timelike one. Also, in this parameter regime, the spacelike surfaces develop an additional turning point in the bulk geometry, corresponding to a non-trivial real solution of $t^{\prime}_{\text{Re}}(r)=\infty$, distinct from the trivial ones at $r=r_0$ and $r=r_h$. In parallel, the timelike extremal surface also develops a turning point, arising from a non-trivial solution of the condition $t^{\prime}_{\text{Im}}(r)=\infty$. Interestingly, the pictorial depiction of the extremal surfaces looks very much similar to the one observed in the scenario $r_h<r_0<r_c$ \cref{surfteewbhreg2}. Despite these intriguing geometric features, the resulting extremal surfaces in the region $r_0 < r_c$ violate the homology condition required for a valid holographic construction of the timelike entanglement entropy. Therefore, in the remainder of this work, we restrict our analysis to the region $r_0 \geq r_c$, where physically admissible solutions are obtained.

\begin{figure*}[h]
    \centering
    \begin{subfigure}[t]{0.33\textwidth}
        \centering
        \includegraphics[width=50mm]{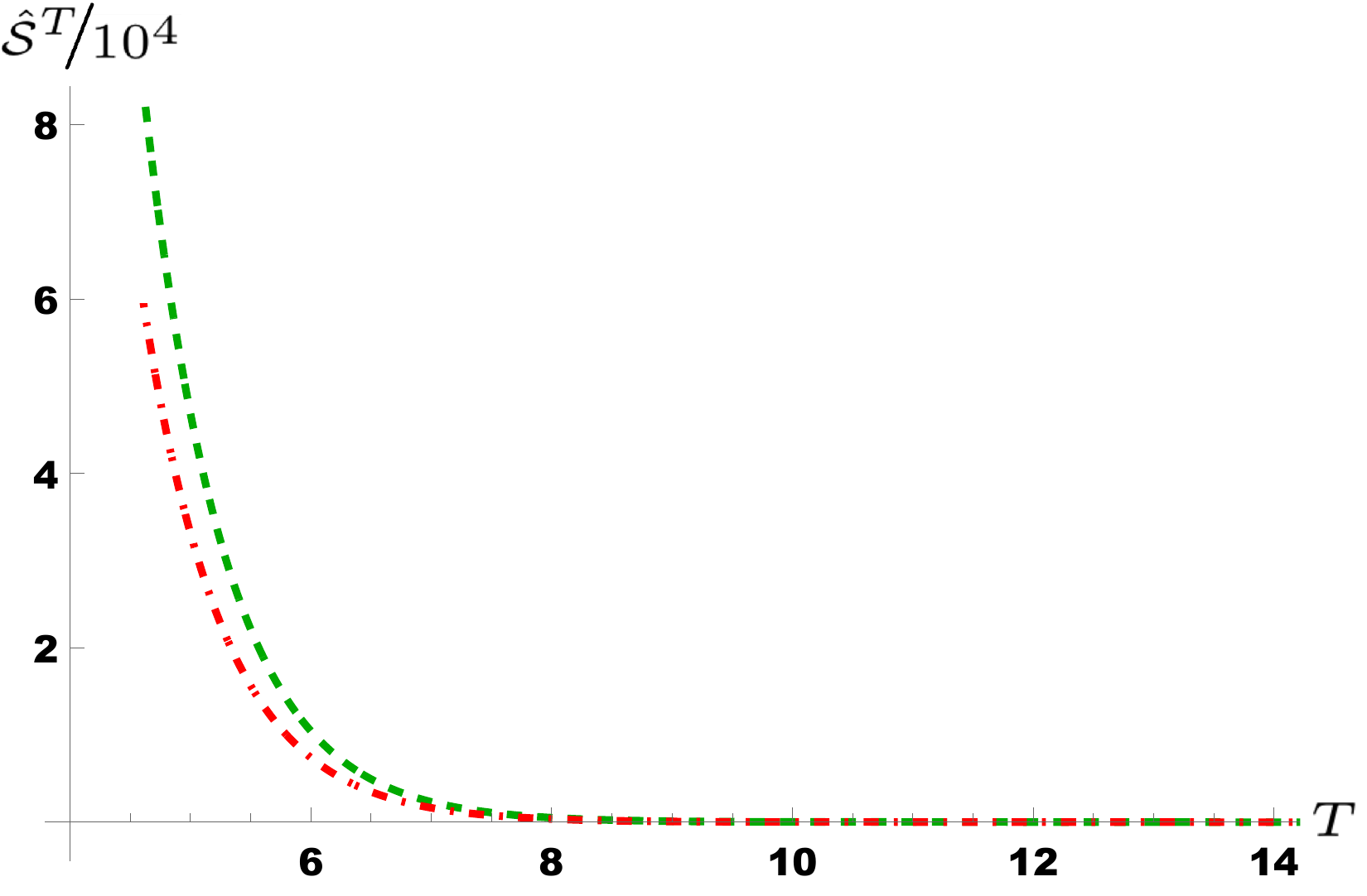}
        \caption{$\frac{\hat{\mathcal{S}}^T}{10^4}$ $vs$ $T$}
        \label{area_tee_bh}
    \end{subfigure}
    \hspace{-1cm}
    \begin{subfigure}[t]{0.33\textwidth}
        \centering
        \includegraphics[width=50mm]{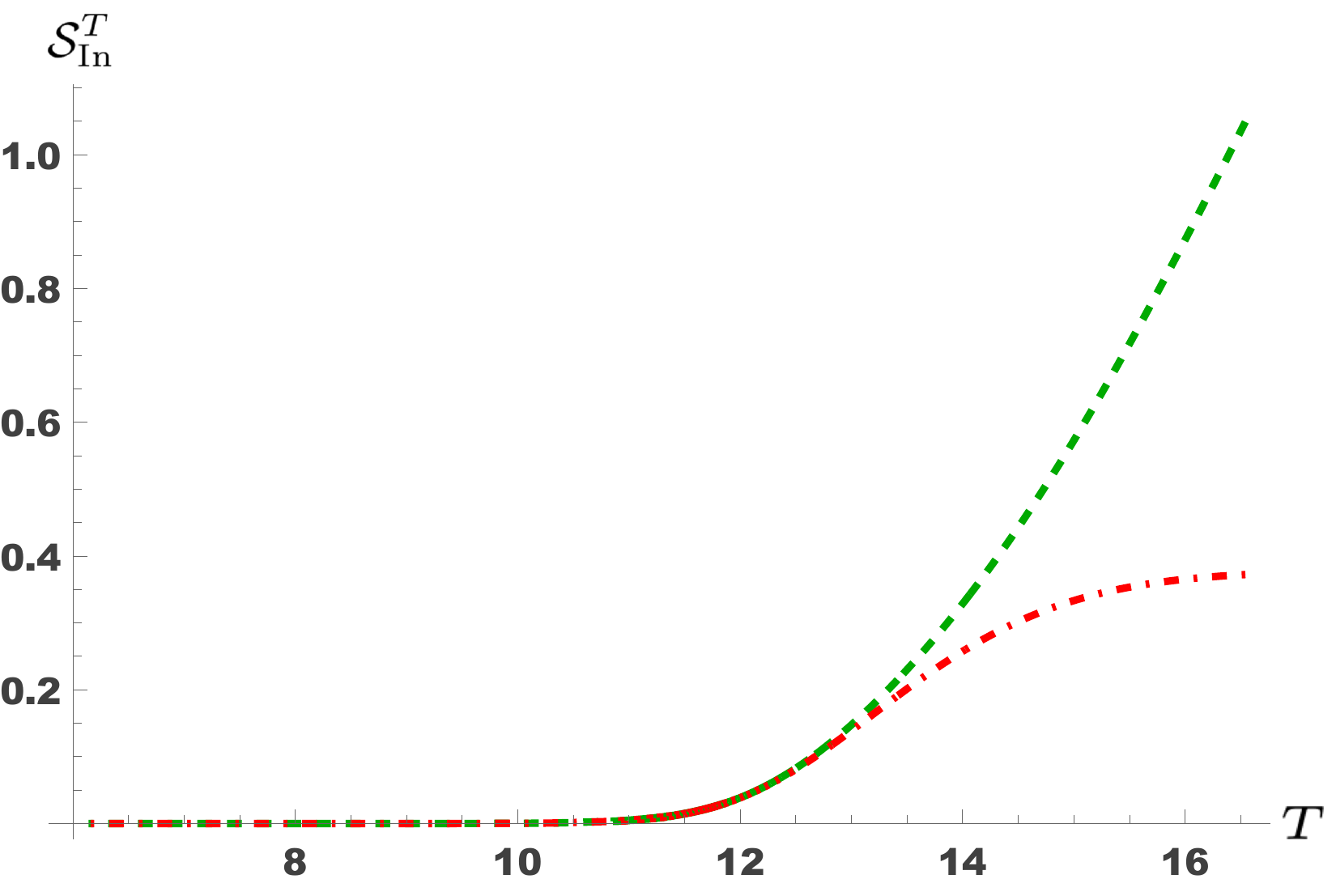}
        \caption{$\mathcal{S}^T_{\text{In}}$ $vs$ $T$}
        \label{area_tee_bh_in}
    \end{subfigure}
    \hspace{-1cm}
    \begin{subfigure}[t]{0.33\textwidth}
        \centering
        \includegraphics[width=50mm]{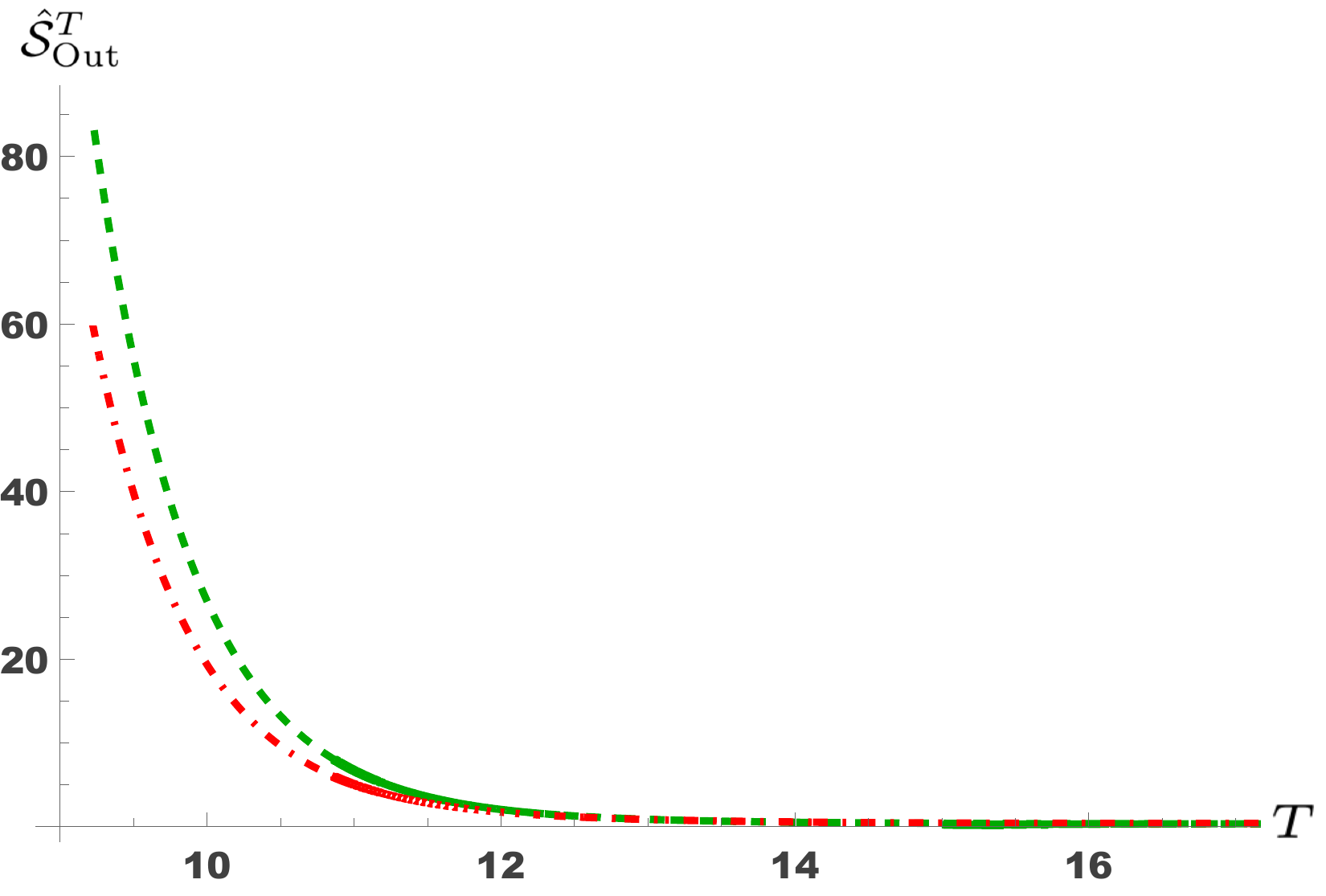}
        \caption{$\hat{\mathcal{S}}^T_{\text{Out}}$ $vs$ $T$}
        \label{area_tee_bh_out}
    \end{subfigure}
    \caption{The areas of the spacelike (green-dashed) and the timelike (red-dot-dashed) surfaces are plotted with increasing boundary subsystem size using \cref{area_int,t_prime_wbh}. Here, we have considered $AdS_4$-Schwarzschild spacetime with the horizon $r_h=1$ and the cutoff $\epsilon=10^{-8}$. (a) Here, we plot the total renormalized areas of the surfaces stretched in both the interior and the exterior of the black hole. The small system size corresponds to high values of areas, whereas it decreases with increasing subsystem size. (b) Here, we plot the black hole interior surface areas without any renormalization. We observe a completely opposite characteristic compared to the first case. The spacelike surface area increases, and the timelike surface area saturates in the region of large subsystem size. (c) We plot the renormalized exterior surface areas where an identical characteristic is observed as the total areas depicted in (a). Interestingly, the values of the exterior surface areas dominate the interior surface areas. 
    }
\end{figure*}

As in the generic spacetime dimension, the analytic evaluation of the boundary length of the subsystem and the areas of the extremal surfaces involve intractable integrations of the integrals given in \cref{t_prime_wbh,area_int}, we delve into a numerical computation. Consequently, the areas of the spacelike and timelike extremal surfaces are computed numerically as functions of the subsystem length (see \cref{area_tee_bh}). The tEE shown in \cref{area_tee_bh} is renormalized by subtracting the contribution of the disconnected surface with $t^\prime(r)=0$, which stretches from the boundary down to the black hole horizon as,
\begin{equation}\label{renorm}
    \hat{\mathcal{S}}^T=\mathcal{S}^T-\mathcal{S}^T_{\text{discon}}~.
\end{equation}
Here we emphasize that, in the renormalization procedure described above, we include only the area contribution of the disconnected solution, $\mathcal{S}^T_{\text{discon}}$ extending from the asymptotic boundary $r\to\infty$ down to the horizon at $r=r_h$. The corresponding contribution is entirely real and removes the divergent part of the real component of the tEE, $\mathcal{S}^T_{\mathrm{Re}}$. However, as discussed previously, the disconnected solution is also extended beyond the horizon, from $r=r_h$ to the curvature singularity at $r=0$. Nevertheless, the area contribution from this interior segment is purely imaginary and a dimension-dependent constant. As a consequence, including this portion of $\mathcal{S}^T_{\text{discon}}$ in the renormalization prescription of \cref{renorm} would amount only to a constant shift in the imaginary part of the tEE, $\mathcal{S}^T_{\mathrm{Im}}$, which does not contain any divergence. Since such a constant shift plays no important role in our numerical analysis, we exclude this contribution of the disconnected solution and restrict $\mathcal{S}^T_{\text{discon}}$ in \cref{renorm} to receive contributions solely from the exterior region of the black hole. The subtraction of the divergent part of $\mathcal{S}^T_{\text{Re}}$, in this way, is necessarily crucial since it substantially improves the numerical stability and provides control over the numerical computations.

The behavior of $\hat{\mathcal{S}}^{T}$ as a function of the subsystem size $T$ reveals interesting features. Particularly, for small subsystem sizes, the timelike surface yields a large area because the corresponding turning point approaches the near-boundary region, where the area contributions become large. Considering the renormalized real part of tEE, we observe that for small subsystem sizes the area of the disconnected surfaces significantly exceeds that of the connected surfaces (\cref{area_tee_bh}). For sufficiently large subsystem sizes, however, a small value of the renormalized area of the spacelike surfaces is obtained. Note that, unlike the standard entanglement entropy, tEE extends at the interior of the black hole which demands an analysis of the surface areas in the interior and the exterior separately. In \cref{area_tee_bh_in}, we plot the areas of the spacelike and the timelike surfaces confined inside the black hole horizon with increasing boundary subsystem size. Note that, unlike \cref{area_tee_bh}, tEE is plotted in \cref{area_tee_bh_in} without any renormalization, as the surface areas do not involve any divergence. Interestingly, here we observe an opposite behavior compared to \cref{area_tee_bh}, as both the surface areas initially show vanishingly small values, and subsequently the area of the spacelike surfaces starts increasing, whereas the area of the timelike surface becomes constant after a brief increase. However, these values are very small, and they are suppressed by the exterior surface areas, which are plotted in \cref{area_tee_bh_out}. Note that the characteristics of the surface areas in \cref{area_tee_bh,area_tee_bh_out} are similar. This also indicates that the connected surface outside the horizon almost coincides with the disconnected geometry, whereas the timelike surface is pushed deeper into the bulk, leading to a reduced area contribution.

We emphasize that the computation of the timelike entanglement entropy in the AdS$_d$–Schwarzschild black hole presented in this section is carried out entirely within the holographic framework. At present, there is no corresponding dual field-theoretic computation of the tEE available in the existing literature that would allow one to directly establish a precise duality between the holographic tEE in generic spacetime dimension and a field-theoretic measure of entanglement along the timelike direction.

\subsection{Holographic tEE}\label{sec_tee_area1}
In this section, we will perform analytic computations of tEE in specific limits. First, we apply a coordinate transformation $z=\frac{1}{r}$ on the metric \cref{sschild_high_d} to have better control on the computation. This transformation alters the holographic direction where the CFT$_d$ lives at $z=0$ and the singularity at $z=\infty$ with the metric,
\begin{align}\label{metric_sch_d1}
   ds_{d+1}^2=\frac{1}{z^2} \left(-f(z)dt^2+\frac{dz^2}{f(z)}+dx_{d-1}^2\right)~.
\end{align}
In the above metric, $f(z)=\left(1-\frac{z^d}{z_h^d} \right)$ is the blackening factor with $z_h$ being the horizon of the geometry. Utilizing the equations of motion in \cref{t_prime_im,t_prime_re}, the subsystem length corresponding to the timelike and the spacelike surfaces can be obtained as,
\begin{align}
   T_{\text{Im}} =2 \int_{z_0}^{\infty}t^\prime_{\text{Im}}(z)dz = 2 \int_{z_0}^{\infty} \frac{dz}{f(z) \sqrt{1-\frac{f(z)}{f\left(z_0\right)}\left(\frac{z_0}{z}\right)^{2 d-2}}}~,\label{interval_int_tEE_im1}\\
   T_{\text{Re}}= 2\int_{0}^{\infty}t^\prime_{\text{Re}}(z)dz = 2\int_{0}^{\infty} \frac{dz}{f(z) \sqrt{1+\frac{f(z)}{f\left(z_0\right)}\left(\frac{z_0}{z}\right)^{2 d-2}}}~.\label{interval_int_tEE_re1}
\end{align}
Following the merging condition $g^{\mu\nu}\partial_\mu \Sigma_{\text{Re}}\partial_\nu \S_{\text{Re}}|_{z\rightarrow \infty}=-g^{\mu\nu}\partial_\mu \Sigma_{\text{Im}}\partial_\nu \S_{\text{Im}}|_{z\rightarrow \infty}$, the extremal surfaces are merged at $z=\infty$. The total subsystem length is then obtained as
\begin{align}\label{interval_tEE1}
   T = T_{\text{Im}} - T_{\text{Re}}~.
\end{align}
Utilizing \cref{area_gen_Im,area_gen_Re}, the extremized area integral for the timelike and spacelike surfaces constituting the tEE can be read for the background \cref{metric_sch_d1} as
\begin{align}\label{area_int_tEE1}
   4 G_N^{(d+1)} \mathcal{S}^T_{\text{Im}} &= 2L^{d-2} \int_{z_0}^{\infty} \frac{dz}{z^{d-1} \sqrt{f(z)} \sqrt{1-\frac{ f \left(z_0\right)}{f (z)}\left(\frac{z}{z_0}\right)^{2 d-2}}}~,\notag\\
   4 G_N^{(d+1)} \mathcal{S}^T_{\text{Re}} &= 2L^{d-2} \int_{\epsilon}^{\infty} \frac{dz}{z^{d-1} \sqrt{f(z)} \sqrt{1+\frac{ f \left(z_0\right)}{f (z)}\left(\frac{z}{z_0}\right)^{2 d-2}}}~,
\end{align}
where $\epsilon$ is the UV cut-off at the asymptotic boundary of AdS black hole geometry. Note that $\mathcal{S}^T_{\text{Re}}$ contains UV divergences since the corresponding spacelike surfaces are stretched towards the asymptotic boundary at $t=\pm T/2$.
The tEE is then expressed as
\begin{align}\label{area_tEE1}
   \mathcal{S}^T = \mathcal{S}^T_{\text{Re}} + \mathcal{S}^T_{\text{Im}}~.
\end{align}
We now introduce the change of variable $y=z/z_0$ under which the tEE can be expressed as
\begin{align}\label{area_int_tEE_y1}
   4 G_N^{(d+1)} \mathcal{S}^T = \frac{2L^{d-2}}{z_0^{d-2}} \Large\Bigg( & \int_{1}^{\infty} \frac{dy}{y^{d-1} \sqrt{f(z_0 y)} \sqrt{1-\frac{ f \left(z_0\right)}{f (z_0y)}y^{2 d-2}}} \notag\\
   +
   &\int_{0}^{\infty} \frac{dy}{y^{d-1} }\left( \frac{1}{\sqrt{f(z_0 y)} \sqrt{1+\frac{ f \left(z_0\right)}{f (z_0y)}y^{2 d-2}}}-1 \right) + \int_{\epsilon/z_0}^{\infty} \frac{dy}{y^{d-1}}\Large\Bigg)~,
\end{align}
where in the second line of the expression, we have added and subtracted appropriate terms to isolate the UV divergence that appears in the real part of the tEE, $\mathcal{S}^T_{\text{Re}}$. Using the same change of variable, the subsystem length $T$ can be expressed as
\begin{align}\label{interval_int_tEE_y1}
   T = 2 z_0 \left( \int_{1}^{\infty} \frac{dy}{f(z_0 y) \sqrt{1-\frac{f(z_0 y)}{f\left(z_0\right)}\frac{1}{y^{2 d-2}}}} -
    \int_{0}^{\infty} \frac{dy}{f(z_0 y) \sqrt{1+\frac{f(z_0 y)}{f\left(z_0\right)}\frac{1}{y^{2 d-2}}}}\right)~.
\end{align}
Substituting \cref{interval_int_tEE_y1} in \cref{area_int_tEE_y1}, we obtain the holographic tEE, $\mathcal{S}^T$ in terms of the subsystem length $T$ and the turning point $z_0$ as 
\begin{align}\label{tEE_y1}
   \mathcal{S}^T = \frac{L^{d-2}}{4 G_N^{(d+1)}}\left( \frac{2}{(d-2) \epsilon^{d-2}} + \frac{T}{z_0^{d-1}} + \frac{2}{z_0^{d-2}}F(z_0)\right)~,
\end{align}
where $F(z_0)$ can be given by
\begin{align}\label{Fz01}
   F(z_0) &= \int _0^{\infty }\frac{dy}{y^{d-1}}\left( \frac{1}{\sqrt{f\left(z_0 y\right)}\sqrt{1+\frac{ f\left(z_0\right)}{f\left(z_0 y\right)}y^{2 d-2}}}\left(1 + \frac{\sqrt{f\left(z_0\right)}}{f\left(z_0 y\right)}y^{2 d-2} \right)-1\right) \notag\\
   & \hspace{1.2cm} + \int _1^{\infty }\frac{dy}{y^{d-1}}\frac{1}{\sqrt{f\left(z_0 y\right)}\sqrt{1-\frac{ f\left(z_0\right)}{f\left(z_0 y\right)}y^{2 d-2}}}\left(1 + i\frac{\sqrt{f\left(z_0\right)}}{f\left(z_0 y\right)}y^{2 d-2} \right)~.
\end{align}
In \cref{tEE_y1}, the divergence part is isolated in the first term, and the other terms depend on the subsystem length $T$, turning point $z_0$ and dimension $d$ of the theory. 

\begin{figure*}[h]
    \centering
    \begin{subfigure}[t]{0.49\textwidth}
        \centering
        \includegraphics[width=73mm]{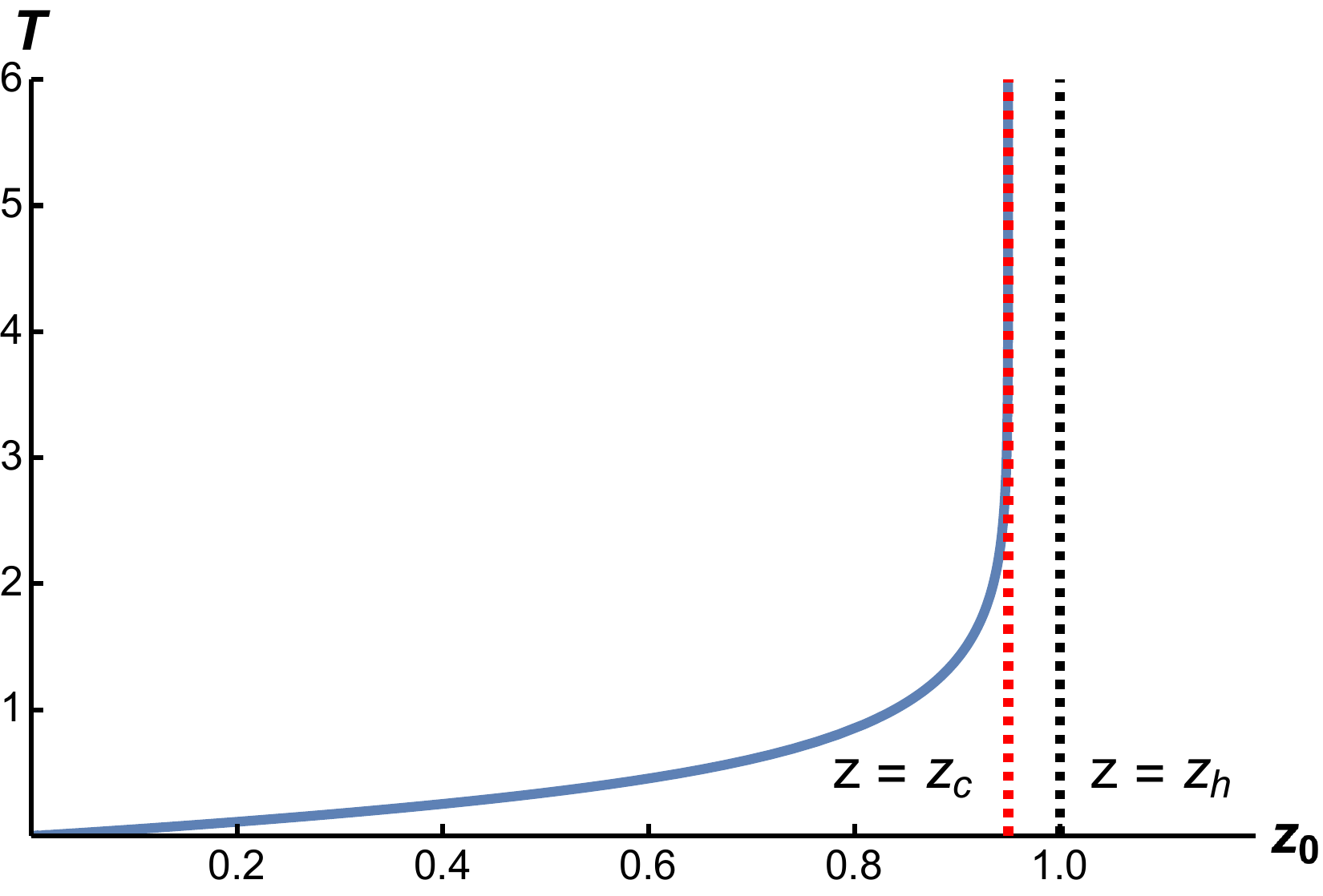}
        \caption{$T$ $vs$ $z_0$}
        \label{t_r01}
    \end{subfigure}
    \hspace{-1cm}
    \begin{subfigure}[t]{0.49\textwidth}
        \centering
        \includegraphics[width=78mm]{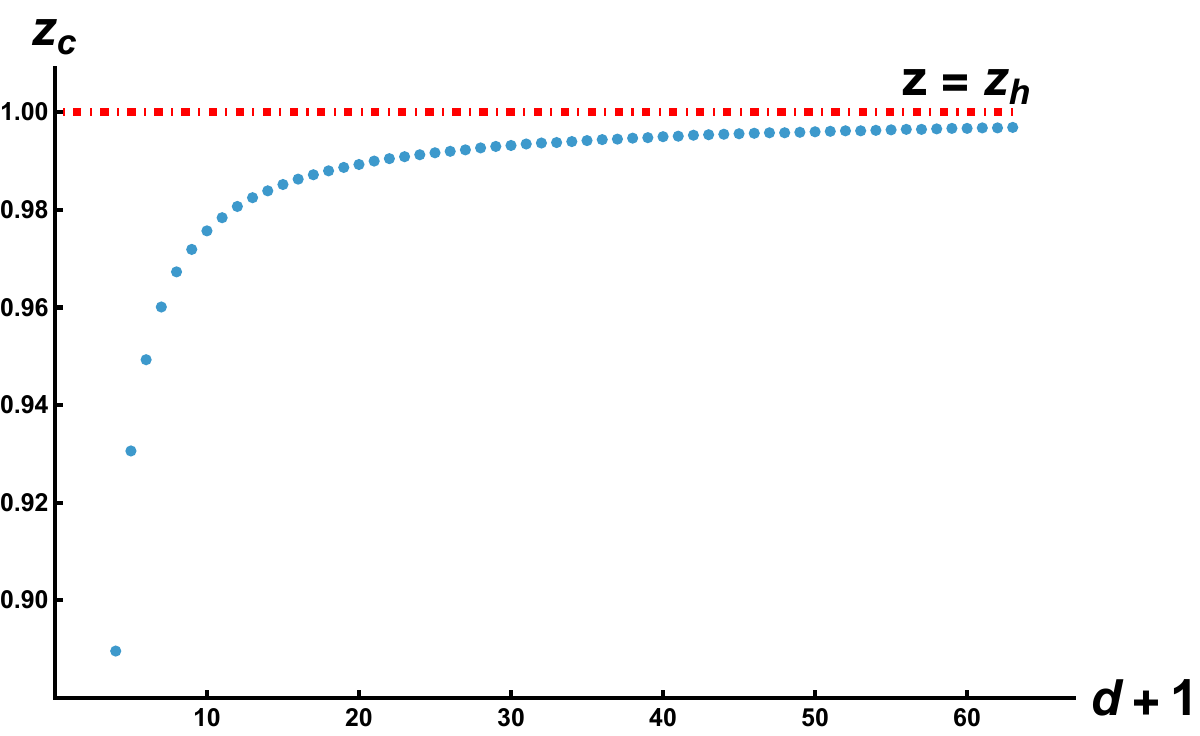}
        \caption{$z_c$ $vs$ $d+1$}
        \label{t_rc1}
    \end{subfigure}
    \caption{Considering the cutoff $\e=10^{-8}$ and the black hole horizon $z_h=1$, we show the behaviour of the subsystem size $T$ with dimension $d$ and the turning point of the timelike surface $z_0$. (a) The total subsystem size $T$ can be read from the turning point of the timelike surface $z_0$. $T$ increases with $z_0$ moving close to the horizon and finally diverges at the critical turning point $z_0=z_c$. For a particular dimension of the AdS space $d$, the critical turning point is fixed. Here we fixed the dimension   of the AdS$_{d+1}$ space to $(5+1)$. (b) The critical turning point $z_c$ moves closer to the horizon $z_h$ with increasing dimension of AdS$_{d+1}$.}
\end{figure*}

Note that, the right hand side of the \cref{interval_int_tEE_y1} depends on $z_0$ and inverting it yields $z_0$ as a function of the total subsystem length $T$. However, computing \cref{interval_int_tEE_y1} analytically for generic $d$ becomes intractable which restricts us to explore \cref{tEE_y1} in detail. Nevertheless we utilize numerical analysis to understand the relation between the turning point $z_0$ and the subsystem length $T$. The plot in \cref{t_r01} shows that, similar to the BTZ black hole case discussed in \cref{sec_tee_btz}, the subsystem length $T$ diverges at a critical value of the turning point $z_0=z_c$. This critical point $z_c$ is not analytically computable because of the involved integration of $t^{\prime}_{\text{Im,Re}}(z)$ arising in the generic dimensional background considered in this section.\footnote{However, one can evaluate the numerical value of the critical turning point $z_c$ by analyzing the roots of $t^{\prime}_{\text{Im,Re}}(z)=\infty$ as described in the previous subsection.} Nevertheless, \cref{t_rc1} demonstrates that as the dimensions of the theory $d$ increase, the critical point $z_c$ moves towards the black hole horizon $z_h$. In the limit $d\rightarrow \infty$, the critical point coincides with the horizon, $z_c=z_h$. Consequently, for large dimensions $d$, the subsystem length $T$ diverges at $z_0=z_c=z_h$. This particular property offers an alternative path to explore the analytic timelike entanglement entropy in the large $d$ and large $T$ regime. 
However, considering only the large $T$ regime where $z_0\to z_c$, the finite part of the tEE is given by
\begin{align}\label{interval_tEE_zc1}
   \frac{4 G_N^{(d+1)}}{L^{d-2}}\mathcal{S}_{finite}^T \simeq  \frac{T}{z_c^{d-1}} + \frac{2}{z_c^{d-2}}F(z_c)~.
\end{align}
In the above expression, $F(z_c)$ is given by \cref{Fz01} with $z_0$ replaced by $z_c$. The first term of the above expression is linear to the subsystem length $T$ with a proportionality constant of $z_c^{1-d}$ for a fixed dimension $d$. This suggests that in the large $T$ limit, the finite part of the holographic tEE takes the form
\begin{align}\label{S_vol_area1}
    \mathcal{S}^T_{finite} \simeq s\,V + \alpha\,A~,
\end{align}
where $s=\frac{1}{4G_N^{(d+1)}z_c^{d-1}}$ and $\alpha=\frac{2}{4G_N^{(d+1)}z_c^{d-2}}F(z_c)$.
Also in the above expression, $V= L^{d-2}T$ is the volume of the strip subsystem $A$ extended in $t$ and $x_{d-2}$ directions and $A=L^{d-2}$ is the area of $\partial A$ along the transverse $x_{d-2}$ directions. Restoring the AdS radius $R$, one can note that $s=\frac{R^{d-1}}{4G_N^{(d+1)}z_c^{d-1}}$ and $\alpha=\frac{2 R^{d-1}}{4G_N^{(d+1)}z_c^{d-2}}F(z_c)$ are constants. Interestingly, the expression in \cref{S_vol_area1} exhibits a structural resemblance to the area law for the subleading area term in the standard entanglement entropy. However, in contrast to entanglement entropy, the timelike entanglement entropy (tEE) includes an intrinsically imaginary contribution which is present only in the area term in \cref{S_vol_area1}. For entanglement entropy, the coefficient of the area term is interpreted as a measure of the effective degrees of freedom of the system along the RG flow, leading to the area theorem \cite{Casini:2012ei,Casini:2016udt,Liu:2012eea,Myers:2010tj,Myers:2010xs,Myers:2012ed}.
On the other hand, existing proofs of the area theorem for entanglement entropy rely fundamentally on the strong subadditivity property \cite{Casini:2012ei}. Since timelike entanglement entropy does not obey strong subadditivity, the formulation and proof of a corresponding timelike area theorem for tEE remains a nontrivial issue. 
Nonetheless, the structure of \cref{S_vol_area1} strongly motivates a deeper investigation into the behavior and interpretation of the area coefficient $\alpha$, particularly in regimes where analytic control is possible.
Specifically, we will examine the large subsystem length limit, $T\to\infty$, as it enables a clear and systematic comparison between the UV and IR coefficients, $\alpha_{\mathrm{UV}}$ and $\alpha_{\mathrm{IR}}$, thereby directly probing their relative magnitudes.

Following \cite{Gushterov:2017vnr}, we first introduce a timelike entanglement density, the difference between the timelike entanglement entropies $\mathcal{S}^T$ and $\mathcal{S}_{0}^T$ of an excited state and a vacuum state of the CFT, respectively, as,
\begin{align}\label{density1}
    \mathcal{S}_{density}^T=\frac{\mathcal{S}^T-\mathcal{S}_{0}^T}{V}~,
\end{align}
where $V$ is the volume of the entangling region at the boundary. The above expression is cut off independent since both $\mathcal{S}^T$ and $\mathcal{S}_{0}^T$ contain the same UV divergences. Focusing on the vacuum state of CFT, $\mathcal{S}_{0}^T$ is computed using  \cref{area_int_tEE1,area_tEE1,interval_int_tEE_im1,interval_int_tEE_re1,interval_tEE1} with the blackening factor $f(z)$ and $f(z_0)$ both set to $1$,
\begin{align}\label{interval_tEE_pure1}
   \mathcal{S}^T_0 = \frac{L^{d-2}}{4 G_N^{(d+1)}}\left( \frac{2}{(d-2) \epsilon^{d-2}} + \frac{T}{z_0^{d-1}} + \frac{2}{z_0^{d-2}}K_1\right)~,
\end{align}
where $K_1$ is a constant given by
\begin{align}\label{K11}
    K_1&=-\int_1^{\infty} \frac{d y}{\sqrt{1-y^{2-2 d}}}\left(1+i y^{2-2 d}\right)
     +\int_0^{\infty} \frac{1}{y^{d-1}}\left(\sqrt{1+y^{2 d-2}}-1\right)~,\notag\\
     &=\sqrt{\pi } \cot \left(\frac{\pi }{2 (d-1)}\right)\frac{\Gamma \left(\frac{d}{2 (d-1)}\right)}{\Gamma \left(\frac{1}{2 (d-1)}\right)}\left( 1-\frac{d-1}{d-2}\sec \left(\frac{\pi }{2 (d-1)}\right)-\frac{i}{d-2}\right).
\end{align}
Furthermore, utilizing \cref{interval_int_tEE_im1,interval_int_tEE_re1,interval_tEE1} with $f(z)=1=f(z_0)$, the subsystem length $T$ can also be obtained analytically for the case of the vacuum state of the CFT as
\begin{align}
    T=z_0 \gamma~,
\end{align}
where $\gamma$ is a numerical constant given by
\begin{align}\label{gamma1}
    \gamma=\sqrt{\pi }\frac{  \Gamma \left(1-\frac{1}{2 (d-1)}\right)}{\Gamma \left(\frac{1}{2}-\frac{1}{2 (d-1)}\right)}\left(1-\sec \left(\frac{\pi }{2 (d-1)}\right)\right)~.
\end{align}
Replacing $z_0$ in terms of the subsystem length $T$ in \cref{interval_tEE_pure1} we get,
\begin{align}\label{interval_tEE_pure_T01}
   \mathcal{S}^T_0 = \frac{L^{d-2}}{4 G_N^{(d+1)}}\left( \frac{2}{(d-2) \epsilon^{d-2}} + \frac{\gamma^{d-1}}{T^{d-2}} + \frac{2 \gamma^{d-2}}{T^{d-2}}K_1\right)~.
\end{align}
Note that the above expression implies that for a large subsystem length $T\to \infty$, the vacuum state timelike entanglement entropy is dominated by the divergent term as
\begin{align}\label{area_tEE_pure_large_dT1}
   \mathcal{S}^T_0 \simeq \frac{L^{d-2}}{4 G_N^{(d+1)}} \frac{2}{(d-2) \epsilon^{d-2}}~.
\end{align}
Accordingly, from \cref{tEE_y1,area_tEE_pure_large_dT1} and restoring the AdS radius $R$, one can observe that in the large subsystem length $T\to \infty$ limit, \cref{density1} reduces to
\begin{align}\label{density_final1}
   \mathcal{S}^T_{density} &\simeq \frac{R^{d-1}}{4 G_N^{(d+1)}}\left( \frac{1}{z_c^{d-1}} + \frac{2}{z_c^{d-2}T}F(z_c)\right).
\end{align}
Note that, we have replaced $z_0$ in \cref{tEE_y1} by $z_c$ as the $T\to\infty$ can only be reached when $z_0\to z_c$. Interestingly, the function $F(z_c)$ in \cref{density_final1} contains both real and imaginary components, which makes the interpretation of $\mathcal{S}^T_{density}$ as a notion of the number of degrees of freedom intrinsically ambiguous. Note that, the complex value of the function $F(z_c)$ also indicates a complex $\alpha$ in \cref{S_vol_area1} which again makes the notion of the area theorem obscure. To find a more realizable probe, we consider 
\begin{align}\label{cfun1}
    \texttt{S}^{T}=\mathcal{S}^T_{Re}+i \,\mathcal{S}^T_{Im}~,
\end{align}
instead of $\mathcal{S}^T$. The quantity $\texttt{S}^{T}$ is motivated by the construction of the holographic timelike $c$-function introduced in \cite{Giataganas:2025div,Giataganas:2025ize}. Note that, in \cref{cfun1}, $\texttt{S}^{T}$ is a real quantity that allows us to propose an area theorem involving the timelike entanglement entropy. Utilizing \cref{cfun1}, we can write,
\begin{align}\label{density_final2}
   \hat{\texttt{S}}^{T}_{density}=\frac{\texttt{S}^{T}_{density}}{s}&=1+ \Delta\alpha \frac{4G_N^{(d+1)}z_c^{d-1}}{R^{d-1}}\frac{1}{T}~,
\end{align}
where $\Delta\alpha$ denotes the difference between the coefficients $\alpha$ appearing in the expression for the tEE, evaluated for the excited state and the vacuum state of the CFT, respectively as \cite{Gushterov:2017vnr}
\begin{align}\label{deltaalpha}
    \Delta\alpha&=\alpha_{IR}-\alpha_{UV}=\frac{2 R^{d-1}}{4G_N^{(d+1)}z_c^{d-2}}\texttt{F}(z_c)~,
\end{align}
with $\texttt{F}(z_c)$ a is a real valued function,
\begin{align}\label{fontedfzc}
   \texttt{F}(z_c) &= \text{Re}[F(z_c)]-\text{Im}[F(z_c)]~.
\end{align}

In principle, the monotonicity can be straightforwardly checked by computing $\hat{\texttt{S}}^{T}_{density}$ upon utilization of \cref{area_int_tEE_y1,interval_int_tEE_y1,interval_tEE_pure_T01} as a function of the subsystem size for a fixed black hole temperature. Note that, for small subsystem size $T\to 0$, $\mathcal{S}^T\sim\mathcal{S}^T_0$ which indicates a vanishing $\hat{\texttt{S}}^{T}_{density}$ whereas for $T\to \infty$ the area term becomes infinitesimally small and as a consequence $\hat{\texttt{S}}^{T}_{density}\to 1$. Interestingly, to confirm monotonicity, it is enough to check if in the large $T$ limit, $\hat{\texttt{S}}^{T}_{density}$ approaches 1 from above or below. The monotonicity is clearly violated or satisfied if $\hat{\texttt{S}}^{T}_{density}\to 1^+$ and $\hat{\texttt{S}}^{T}_{density}\to 1^-$ respectively, in the large subsystem limit. The analysis therefore reduces to determining the signature of the coefficient $\Delta\alpha$, with $\Delta\alpha<0$ indicating a well-behaved timelike area theorem and $\Delta\alpha>0$ signaling its violation. Unfortunately, the computation of $\Delta\alpha$, even for the large subsystem size limit, becomes intractable as the analytic form of $z_c$ is unknown. However, a large-$d$ consideration along with $T\to\infty$ makes the computation feasible as it yields $z_0=z_c=z_h$. Consequently, the finite part of the holographic tEE described in \cref{interval_tEE_zc1} takes the form
\begin{align}\label{interval_tEE_zh1}
   \frac{4 G_N^{(d+1)}}{L^{d-2}}\mathcal{S}_{finite}^T \simeq  \frac{T}{z_h^{d-1}} + \frac{2}{z_h^{d-2}}F(z_h)~,
\end{align}
where, $F(z_h)$ is expressed as,
\begin{align}\label{Fzh_large_d1}
    F(z_h) &= \int_0^{\infty}\frac{dy}{y^{d-1}}\left( \frac{1}{\sqrt{1-y^d}}-1\right) + \int_1^{\infty}\frac{dy}{y^{d-1}} \frac{1}{\sqrt{1-y^d}}\notag\\
    &=\int_0^{1}\frac{dy}{y^{d-1}} \left( \frac{1}{\sqrt{1-y^d}}-1\right)-\int_1^{\infty}\frac{dy}{y^{d-1}} + 2\int_1^{\infty}\frac{dy}{y^{d-1}} \frac{1}{\sqrt{1-y^d}}\notag\\
    & =\frac{y^{2-d}}{d-2} \left(1- \, _2F_1\left[\frac{1}{2},\frac{2}{d}-1,\frac{2}{d},y^d\right]\right)\Bigg{|}_{y=0}^{y=1} +\frac{y^{2-d}}{d-2}\Bigg{|}_{y=1}^{y=\infty} - \frac{2~y^{2-d}}{d-2} \,_2F_1\left[\frac{1}{2},\frac{2}{d}-1,\frac{2}{d},y^d\right]\Bigg{|}_{y=1}^{y=\infty}\notag\\
    & \simeq \frac{1}{4} + \frac{ 2\log 2 -1}{2 d}-\frac{ i \pi}{ d}+\mathcal{O}\left(\frac{1}{d^2}\right)~.
\end{align}
Note that, $F(z_h)$ contains both real and imaginary parts, where the imaginary part only appears in $\frac{1}{d}$ order. Also, in this limit, the vacuum state tEE of the CFT is still given by \cref{area_tEE_pure_large_dT1} since at large dimensions $d$, the constants $K_1$ and $\gamma$ from \cref{K11,gamma1} behave as
\begin{align}\label{K1_gamma}
    K_1 \simeq -\frac{1+i}{d}~, \hspace{1cm} \gamma \simeq -\frac{\pi ^2}{8 d^2}~.
\end{align}
Utilizing \cref{fontedfzc}, we compute
\begin{equation}\label{fontedfzh2}
    \texttt{F}(z_h)=\frac{1}{4} + \frac{ 2\log 2 +2\pi -1}{2 d}~,
\end{equation}
which is a positive quantity that also makes $\Delta\alpha$ positive in \cref{deltaalpha}. 
Utilizing \cref{fontedfzh2}, the above expression implies
\begin{align}
    \alpha_{UV}&=\alpha_{IR}-s\, z_h \left(\frac{1}{2}+ \frac{ 2\log 2 +2\pi -1}{d} \right)\label{areathtee1}~,
\end{align}
where the term inside the parenthesis is always positive. It yields that in the large-$d$ regime, $\alpha_{IR}>\alpha_{UV}$ which clearly violates the timelike area theorem.

So far, we have focused on the geometric construction of the extremal surfaces and utilized them to compute the holographic tEE in different black hole backgrounds. In the following section, we shift our attention to a detailed analysis of the behavior of these extremal surfaces in the near-horizon region. As we shall demonstrate, the near-horizon limit is of particular importance, as universal aspects of black hole dynamics are most naturally realized in this regime.

\section{Near-horizon behaviour of the surfaces}\label{sec_tee_growth}
In this section we investigate the universal near-horizon behaviour of the bulk extremal surfaces that constitute the holographic timelike entanglement entropy (tEE). Our aim is to understand how the spacelike and the timelike branches of the holographic tEE behave as they approach the black-hole horizon and determine their characteristic growth rates at late times. This analysis parallels the well-known near-horizon behaviour of Ryu–Takayanagi (RT) surfaces in entanglement entropy, which encodes hallmark signatures of quantum chaos and fast scrambling. We will show that an analogous structure emerges for the extremal surfaces associated with the holographic tEE. 

\subsection*{BTZ black hole}
We begin our analysis by considering the AdS$_3$ BTZ black hole metric described in \cref{btz_geo},
\begin{align}
    ds^2=-\left(r^2-r_h^2\right)dt^2+\frac{dr^2}{r^2-r_h^2}+r^2 d\phi^2~.
\end{align}
and analyze the behaviour of the bulk extremal surfaces of the holographic tEE. The asymptotic boundary of this geometry is at $r\to\infty$ where the dual CFT$_2$ is located with a temporal subsystem $A \equiv -T/2 \leq t \leq T/2$. We are particularly interested in the region where the bulk extremal surfaces corresponding to the tEE of the subsystem $A$ approach and asymptotically wrap the event horizon $r_h$. In this limit, the surfaces probe the universal Rindler structure characteristic of black-hole horizons, enabling a precise characterization of their late-time behaviour and exponential approach governed by the surface gravity. To capture this regime, we parameterize the bulk profiles as
\begin{equation}\label{near_rh}
    r_{\text{Im,Re}}(t)=r_h \left(1+\epsilon~  y_{\text{Im,Re}}(t)^2\right)~.
\end{equation}
In the above parametrization, we have considered the near-horizon profile of the extremal surfaces just outside the black hole. One may likewise consider the corresponding near-horizon profile inside the black hole, where an analogous analysis, as described below, can be carried out to obtain the exact results.
In \cref{near_rh}, $\epsilon$ is an arbitrarily small number which keeps the surface close to the horizon $r_h$ and the function $y_{\text{Re,Im}}(t)\to 0$ monotonically as $t\to \infty$ where the bulk surfaces wrap the horizon asymptotically. We now consider an ansatz to the solutions of the bulk profiles of the extremal surfaces as \cite{Dong:2022ucb,Mezei:2016wfz},
\begin{equation}\label{btz_ansatz}
    y_{\text{Im,Re}}(t)\sim y_0\,e^{-\lambda_{\text{Im,Re}} t}~,
\end{equation}
where $\lambda_{\text{Im,Re}}$ measures the exponential rate at which the surfaces approach the horizon, and $y_0$ is a numerical constant. Utilizing the near-horizon expansion of \cref{near_rh} and substituting it into the equations of motion described in \cref{t_prime_im,t_prime_re}, the extremal surface dynamics in the vicinity of the black-hole horizon $r_h$ reduce, at leading order in $\epsilon$, to the simple relation
\begin{align}
    y_{\text{Im,Re}}'(t)^2-r_h^2\, y_{\text{Im,Re}}(t)^2=0~.
\end{align}
Furthermore, substituting the ansatz considered in \cref{btz_ansatz}, we obtain identical growth rates for both the timelike and spacelike surfaces as
\begin{equation}
    \lambda_{\text{Im,Re}}=\frac{2\pi}{\beta}~,
\end{equation}
where $\beta$ is the inverse Hawking temperature of the BTZ black hole. This behaviour admits a clear physical interpretation, implying that near the BTZ horizon, the gravitational redshift forces the extremal surfaces to fall toward the horizon in an exponentially slow manner. Both the spacelike and timelike branches of the holographic tEE exhibit this universal approach, with a decay rate precisely set by the horizon temperature of the black hole, even though the extremal surfaces are geometrically distinct. Moreover, the value matches precisely the Maldacena–Shenker–Stanford (MSS) bound $\lambda_L^{BTZ}=\frac{2\pi}{\beta}$ on the Lyapunov exponent for quantum chaos. This universal exponential behaviour provides a compelling geometric manifestation of fast scrambling in the holographic tEE framework.

\subsection*{Black holes with hyperscaling violation}
Building upon the BTZ analysis, we now extend the near-horizon study of the extremal surfaces to a broader class of geometries exhibiting hyperscaling violation. Such backgrounds arise naturally in holographic theories with nontrivial scaling properties and capture effective IR dynamics of strongly coupled systems with reduced dimensionality. The finite-temperature deformation of these theories is described by the general metric
\cite{Iizuka:2011hg,Swingle:2011np,Dong:2012se}
\begin{equation}
    ds_{d+1}^2=\left(\frac{z}{z_F}\right)^{\frac{2 \theta }{d-1}}z^{-2} \left(-z^{-2 (\xi-1)}f(z)dt^2 +\frac{dz^2}{f(z)}+dx_{d-1}^2\right)~,
\end{equation}
where the dynamical critical exponent $\xi$ and hyperscaling violation parameter $\theta$ \cite{Fisher:1986zz} control the anisotropic and scale-covariant structure of the dual field theory. The geometry is supported by the blackening factor and the associated black-hole temperature, given respectively by
\begin{equation}\label{fz_beta_hyp}
    f(z)=1-\left(\frac{z}{z_h}\right)^{d+\xi-\theta-1}, ~~~~~~~~\beta_{\text{Hyp}}=\frac{4\pi z_h^\xi}{d+\xi-\theta-1}~.
\end{equation}
Here $z_F<z_h$ is considered to study finite temperature effects on a regime with hyperscaling violation. Note that the asymptotic boundary of the metric is at $z=0$ where the dual CFT$_d$ is located. We consider a strip-like subsystem $A \equiv -T/2 \leq t \leq T/2$ in the CFT$_d$ in the Lorentzian time direction $t$ and on a constant $x_i$ slice. The corresponding extremized area integral can be read from \cref{area_gen} as
\begin{align}\label{area_int_hyp}
   4 G_N^{(d+1)} \mathcal{S} = L^{d-2} \int dz~ z^{-(d-\theta -1)} z_F^{-\theta }\sqrt{\frac{1}{f(z)}-z^{-2 (\xi-1)} f(z) t'(z)^2}~,
\end{align}
with $L^{d-2}$ denoting the transverse spatial volume in the $x_{d-2}$ directions and $G_N^{(d+1)}$ the Newton constant of the $(d+1)$-dimensional AdS geometry.
The holographic dual surfaces of the tEE lie on a hypersurface $dx_i=0$ with an induced metric
\begin{equation}
    \gamma _{\alpha \beta } dx^{\alpha } dx^{\beta }=\left(\frac{z}{z_F}\right)^{\frac{2 \theta }{d-1}}z^{-2} \left(dx_{d-2}^2+dz^2 \left(\frac{1}{f(z)}-\frac{z^{-2 (\xi-1)}f(z) }{z'(t)^2}\right)\right). 
\end{equation}
which parallels the BTZ structure described above but now encodes the nontrivial scaling through $(\theta,\xi)$. The equations of motion for these bulk hypersurfaces can be obtained from \cref{area_int_hyp} as
\begin{align}
    {t^{\prime}_{\text{Im}}}(z)^2 &= \frac{z^{2 (\xi -1)}}{f(z)^2 \left(1-\frac{f(z)}{f\left(z_0\right)}\left(\frac{z}{z_0}\right)^{-2 (d +\xi -\theta-2)}\right)}~,\label{t_prime_hyp_im}\\
    {t^{\prime}_{\text{Re}}}(z)^2 &=\frac{z^{2 (\xi -1)}}{f(z)^2 \left(1+\frac{f(z)}{f\left(z_0\right)}\left(\frac{z}{z_0}\right)^{-2 (d +\xi -\theta-2)}\right)}~.\label{t_prime_hyp_re}
\end{align}
As earlier, we focus on the late-time behavior, where the bulk extremal surfaces asymptotically approach the horizon $z_h$ from the exterior. We parameterize this regime by
\begin{equation}\label{nearhorhyp}
    z(t)=z_h \left(1-\epsilon \, y_{\text{Im,Re}}(t)^2\right)~,
\end{equation}
which isolates the universal Rindler region near the black hole horizon. 
Similar to the BTZ black hole scenario, one may again consider the near-horizon profiles of extremal surfaces situated just inside the black hole, and applying the same methodology as in the exterior analysis then yields an exact result.
Now, following this near-horizon solution in \cref{nearhorhyp}, the leading expansion of the blackening factor $f(z)$ can be approximated as
\begin{equation}
    f(z)\approx \epsilon  y(t)^2 (d+\xi-\theta-1)=(d+\xi-\theta-1) \left(1-\frac{z(t)}{z_h}\right)~.
\end{equation}
Finally, upon substituting the near-horizon expansion of \cref{nearhorhyp} into the equations of motion described in \cref{t_prime_hyp_im,t_prime_hyp_re}, the dynamics of the bulk extremal surfaces simplify considerably. In the leading order of the near-horizon parameter $\epsilon$, the equations governing both the timelike and spacelike branches reduce to
\begin{equation}\label{EOM_hyp}
    4 r_h^{2 \xi }y_{\text{Im,Re}}'(t)^2-(d +\xi-\theta-1 )^2 y_{\text{Im,Re}}(t)^2=0~.
\end{equation}
Considering a similar ansatz as \cref{btz_ansatz}, the solutions for the growths corresponding to both the timelike and spacelike surfaces are obtained as,
\begin{align}
    \lambda_{\text{Im,Re}} &=\frac{1}{2}z_h^{-\xi}(d+\xi-\theta-1 )=\frac{2\pi}{\beta_{\text{Hyp}}}~,
\end{align}
where in the last equality we have utilized \cref{fz_beta_hyp}. 
Thus, despite the presence of nontrivial scaling exponents, the black hole horizon enforces a universal exponential fall-off for both tEE branches, governed solely by the Hawking temperature $\beta_{\text{Hyp}}$. Instead, the gravitational redshift in the universal Rindler region forces both the bulk surfaces to obey $y(t) \sim e^{2\pi t/\beta_{\text{Hyp}}}$. This is precisely the same MSS-saturating exponential rate encountered in the BTZ geometry, now generalized to black holes with arbitrary $\theta$ and $\xi$. The key difference is that the combination $d+\xi-\theta-1$ plays the role of an effective IR scaling dimension, replacing the pure AdS value.

\section{Discussions}\label{dis}
The present work has been devoted to exploring the holographic construction of timelike entanglement entropy (tEE) in black hole geometries, extending the formalism introduced in earlier studies \cite{Basak:2023otu,Afrasiar:2024lsi}. The analysis begins with the BTZ black hole background, chosen for its analytical simplicity and its ability to capture the essential geometric features relevant to holographic entanglement. Using the Lorentzian signature of the BTZ metric, we examine the structure of the extremal surfaces associated with the holographic tEE. In this framework, the holographic tEE includes contributions from a pair of spacelike surfaces stretching between the asymptotic boundary and the singularity, and a timelike surface that lies entirely within the bulk, extending from a turning point $r_0$ to the black hole singularity. From the analysis of the equations of motion, we observe that these extremal surfaces, unlike the conventional Ryu–Takayanagi (RT) surface for standard entanglement entropy, are not confined to the exterior region of the black hole. Instead, they naturally extend across the horizon, exploring both the exterior and interior domains of spacetime.
Remarkably, as the surfaces approach the horizon, their slopes diverge symmetrically from both sides, satisfying $t^{\prime}_{\text{Re}}(r_h)=t^{\prime}_{\text{Im}}(r_h)=\pm\infty$. This identical asymptotic characteristics indicates that the real and imaginary branches of the holographic tEE exhibit the same near-horizon geometric behavior, reflecting a common underlying structure in their extremal embeddings.

A particularly novel aspect of this construction is that the timelike surface persists in both the interior and the exterior regions of the black hole horizon, which departs from the formulations presented in earlier literature. Computing the corresponding areas of the extremal surfaces and the subsystem length at the asymptotic boundary purely from holographic principles, we found that our results precisely reproduce those obtained earlier from field-theoretic analyses, thereby providing a strong consistency check for our formalism of the holographic tEE.
In this context, the subsystem length $T$ at the asymptotic boundary depends on the turning point
$r_0$, with $T$ increasing as $r_0$ is pushed deeper into the bulk. However, the analysis indicates the existence of a critical turning point $r_0=r_c$ at which the subsystem length $T$ diverges. Beyond this critical value, the turning point cannot be extended further, as the associated subsystem size ceases to be physically well defined.

Building upon the insights gained from the BTZ geometry, we generalize the construction to higher dimensions by considering the AdS$_{d+1}$ Schwarzschild black hole, which captures the essential features of holographic tEE in generic spacetime dimensions. The underlying principles in this geometry remain similar to the BTZ black hole scenario. The holographic tEE arises from extremizing the combined contributions of spacelike and timelike surfaces, but the complexity of the corresponding integrals grows rapidly with the number of dimensions. While the analytic computation of the holographic tEE becomes intractable, the structure lends itself well to numerical evaluations for specific choices of the parameters. In this setting, the numerically determined areas of the extremal surfaces yield the renormalized quantity $\hat{\mathcal{S}}^{T}=\mathcal{S}^{T}-\mathcal{S}^{T}_{\text{discon}}$, where only the divergent real contribution is subtracted by $\mathcal{S}^{T}_{\text{discon}}$ and the imaginary finite part remains unaffected. The resulting behavior shows that, for sufficiently small boundary subsystems, the turning point of the timelike extremal surface is driven toward the UV region, leading to a large area contribution. In contrast, the corresponding spacelike surfaces develop a small area contribution in this limit, causing the renormalized real part $|\hat{\mathcal{S}}^{T}_{\text{Re}}|$ to be large as the subsystem size approaches zero. However, for large boundary subsystems, the exterior spacelike surfaces become nearly indistinguishable from the disconnected surfaces outside the horizon, while the timelike surface along with its associated turning point is located deeper into the bulk. This behavior drives both the renormalized real component $\hat{\mathcal{S}}^{T}_{\text{Re}}$ and the corresponding imaginary component $\mathcal{S}^{T}_{\text{Im}}$ of the tEE to vanish in the large-subsystem regime. We also examine how the areas of the spacelike and timelike surfaces located inside the black hole horizon vary with increasing boundary subsystem size. Unlike the previously discussed exterior case, no renormalization is required here because the interior surfaces do not contain divergent contributions. Interestingly, the qualitative behaviors are reversed, where both surface areas begin with nearly vanishing values, after which the spacelike surface area gradually increases, while the timelike surface area shows a brief rise before settling to an approximately constant value. Overall, these interior contributions remain very small and are dominated by the corresponding exterior surface areas.
The numerical analysis also reveals a behavior analogous to the BTZ black hole case, indicating the existence of a critical turning point $r_0=r_c$ in the bulk. When $r_0$ is moved towards this critical value, $r_c$ the subsystem length at the asymptotic boundary diverges, and beyond this point, the subsystem ceases to exist as a well-defined real quantity. Interestingly, it is observed that the critical point $r_c$ steadily approaches the horizon of the black hole $r_h$ with an increasing number of the dimensions $d$ of the theory. Additionally, within this framework of the AdS$_d$ Schwarzschild black hole, the spacelike and the timelike surfaces, denoted by $\S_{\text{Re,Im}}$, can now be merged smoothly at the black hole singularity, $r=0$ satisfying the condition $g^{\m\n}\pp_\m \S_{\text{Re}}\pp_\n \S_{\text{Re}}|_{r=0}=g^{\m\n}\pp_\m \S_{\text{Im}}\pp_\n \S_{\text{Im}}|_{r=0}=0$. This indicates that the extremal surfaces momentarily become null-like at $r=0$ in contrast to the behavior observed in the BTZ black hole case. 

Extending this geometric analysis, we further investigate this background to examine the timelike area theorem for the holographic tEE in two significant physical limits, specifically at large subsystem length and large spacetime dimensions. In the large subsystem length limit, the finite part of the holographic tEE scales with the volume at leading order, with subleading correction going with the area of the strip subsystem considered at the asymptotic boundary. In this context, we compute the timelike entanglement density and study its monotonicity which represents if a timelike area theorem can be satisfied or violated. Within this framework, the coefficient of the area term in the expression of the timelike entanglement density in the large subsystem size limit emerges as a crucial indicator of the timelike area theorem. We show that, in the large $d$ and large subsystem size limit, this coefficient approaches a positive constant value at leading orders, which indicates a violation of the timelike area theorem.

In continuation of our analysis, we have further explored the near-horizon dynamics of bulk extremal surfaces corresponding to the timelike entanglement entropy (tEE) and compared their behavior with that of the usual spacelike Ryu–Takayanagi (RT) surfaces. Beginning with the BTZ black hole, we found that both the spacelike and timelike surfaces exhibit an identical exponential growth characterized by the Lyapunov exponent saturating the Maldacena–Shenker–Stanford (MSS) bound, $\lambda_{\text{Re,Im}} = 2\pi/\beta$. This result reflects the universality of chaotic information spreading in $(2+1)$-dimensional AdS spacetimes. Extending the analysis to bulk geometries with Lifshitz scaling and hyperscaling violation, we observed a similar growth $\lambda_{\text{Re,Im}} = 2\pi/\beta_{\text{Hyp}}$ with the black hole inverse temperature, $\beta_{\text{Hyp}}$ which depends on the background exponents $(\xi, \theta, d)$. Interestingly, the growth saturates the MSS bound. Note that, although the timelike and the spacelike surfaces are geometrically distinct, the near boundary growth for both the surfaces show a similar characteristics. Our results therefore suggest that, the tEE framework offers an intriguing geometric extension of entanglement concepts into timelike directions which also characterizes the chaos and information scrambling near black hole horizons.

\section*{Acknowledgment} 
We thank Dimitrios Giataganas for the collaboration in the preliminary stage of this work. We thank Dimitrios Giataganas, Matti Jarvinen, Niko Jokela, Carlos Nunez, Dibakar Roychowdhury and Tadashi Takayanagi for useful comments on the draft. We also thank Banashree Baishya and Adrita Chakraborty for useful discussions.
The research work of MA is supported by NSFC, China (Grant No. 12275166 and No. 12311540141). 
The research work of JKB is supported by the Brain Pool program funded by the Ministry of Science and ICT through the National Research Foundation of Korea (RS-2024-00445164).
This work was supported by the Basic Science Research Program through the National Research Foundation of Korea (NRF) funded by the Ministry of Science, ICT \& Future Planning (NRF-2021R1A2C1006791), the framework of international cooperation program managed by the NRF of Korea (RS-2025-02307394), the Creation of the Quantum Information Science R\&D Ecosystem (Grant No. RS-2023-NR068116) through the National Research Foundation of Korea (NRF) funded by the Korean government (Ministry of Science and ICT), the Gwangju Institute of Science and Technology (GIST) research fund (Future leading Specialized Resarch Project, 2025) and the Al-based GIST Research Scientist Project grant funded by the GIST in 2025. This research was also supported by the Regional Innovation System \& Education(RISE) program through the (Gwangju RISE Center), funded by the Ministry of Education(MOE) and the (Gwangju Metropolitan City), Republic of Korea(2025-RISE-05-001).

\bibliographystyle{JHEP}

\bibliography{timelike}

\end{document}

%% file: no-revtex.tex
\newcommand{\preprint}[1]{\begin{table}[t]  
             \begin{flushright}               
             {#1}                             
             \end{flushright}                 
             \end{table}}                     
\renewcommand{\title}[1]{\vbox{\center\LARGE{#1}}\vspace{5mm}}
\renewcommand{\author}[1]{\vbox{\center#1}\vspace{5mm}}
\newcommand{\address}[1]{\vbox{\center\em#1}}
\newcommand{\email}[1]{\vbox{\center\tt#1}\vspace{5mm}}

\renewcommand{\date}[1]{\vbox{\center#1}}



%% file: jm-tex-macros-public.tex
\usepackage{amssymb}
\usepackage{amsmath,bm}
\usepackage{amssymb}
\usepackage{graphicx}
\usepackage{amsfonts}         
\usepackage{fancybox}   

\usepackage{enumitem}

\usepackage{slashed}

\usepackage{epstopdf}
\usepackage[usenames,dvipsnames]{xcolor}
\usepackage[pagebackref, bookmarks={false}, pdfauthor={JD}, pdftitle={Aamra_Jethay_Mori_Ghure}]{hyperref}
\hypersetup{colorlinks=true, linkcolor=darkblue, citecolor=red, filecolor=OliveGreen, urlcolor=blue, filebordercolor={.8 .8 1}, urlbordercolor={.8 .8 0}}
\usepackage{soul}
\setstcolor{Red}

\usepackage{wrapfig}

\definecolor{jazzberryjam}{rgb}{0.65, 0.04, 0.37}
\definecolor{lust}{rgb}{0.9, 0.13, 0.13}
\definecolor{sandybrown}{rgb}{0.96, 0.64, 0.38}
\definecolor{mountainmeadow}{rgb}{0.19, 0.73, 0.56}
\definecolor{glaucous}{rgb}{0.38, 0.51, 0.71}
\definecolor{chromeyellow}{rgb}{1.0, 0.65, 0.0}
\definecolor{emerald}{rgb}{0.31, 0.78, 0.47}
\definecolor{deepsaffron}{rgb}{1.0, 0.6, 0.2}
\definecolor{darkgreen}{rgb}{0,0.4,0}
\definecolor{darkred}{rgb}{0.4,0,0}
\definecolor{darkblue}{rgb}{0,0,0.4}
\definecolor{lightblue}{rgb}{.6,.6,0.9}

\definecolor{uglybrown}{rgb}{0.8,  0.7,  0.5}

\definecolor{palatinatepurple}{rgb}{0.41, 0.16, 0.38}
\definecolor{celebrationcolor}{rgb}{0.75,  0.0,  0.9}

 \usepackage{mdframed}

\usepackage{framed}
\definecolor{shadecolor}{rgb}{0.90,0.90,0.90}

\usepackage{tkz-euclide}

\usepackage{makeidx}
\usepackage{cite}
\usepackage{bm}
\usepackage{geometry}
\usepackage{braket}
\usepackage[margin=20pt,small]{caption}

\usepackage[toc]{appendix}

\usepackage{tikz}
\usetikzlibrary{calc,decorations.markings,arrows.meta,shapes.misc,decorations.pathmorphing,calc,bending}

\usetikzlibrary{arrows.meta,shapes.misc,decorations.pathmorphing,calc,bending}

\tikzset{
  branch point/.style={cross out,draw=black,fill=none,minimum size=2*(#1-\pgflinewidth),inner sep=0pt,outer sep=0pt}, 
  branch point/.default=5
}
\tikzset{
  branch cut/.style={
    decorate,decoration=snake,
    to path={
      (\tikztostart) -- (\tikztotarget) \tikztonodes
    },
    }
  }

\input epsf

\newlength{\extraspace}
\newlength{\extraspaces}
\def\be{\begin{equation}}
\def\ee{\end{equation}}

\newcommand{\bea}{\begin{eqnarray}}
\newcommand{\eea}{\end{eqnarray}}

\def\II{\relax{I\kern-.10em I}}

\def\IB{\relax{\rm I\kern-.18em B}}

\def\ID{\relax{\rm I\kern-.18em D}}
\def\IE{\relax{\rm I\kern-.18em E}}
\def\IF{\relax{\rm I\kern-.18em F}}
\def\IG{\relax\hbox{$\inbar\kern-.3em{\rm G}$}}
\def\IGa{\relax\hbox{${\rm I}\kern-.18em\Gamma$}}
\def\IH{\relax{\rm I\kern-.18em H}}
\def\II{\relax{\rm I\kern-.18em I}}
\def\IK{\relax{\rm I\kern-.18em K}}

\def\inbar{\,\vrule height1.5ex width.4pt depth0pt}

\def\IR{\mathbb{R}}

\def\lp10{\ell_p^{10}}
\def\lp11{\ell_p^{11}}
\def\R11{R_{11}}

\def\frac#1#2{{#1 \over #2}}

\newdimen\tableauside\tableauside=1.0ex
\newdimen\tableaurule\tableaurule=0.4pt
\newdimen\tableaustep
\def\phantomhrule#1{\hbox{\vbox to0pt{\hrule height\tableaurule width#1\vss}}}
\def\phantomvrule#1{\vbox{\hbox to0pt{\vrule width\tableaurule height#1\hss}}}
\def\sqr{\vbox{%
  \phantomhrule\tableaustep
  \hbox{\phantomvrule\tableaustep\kern\tableaustep\phantomvrule\tableaustep}%
  \hbox{\vbox{\phantomhrule\tableauside}\kern-\tableaurule}}}
\def\squares#1{\hbox{\count0=#1\noindent\loop\sqr
  \advance\count0 by-1 \ifnum\count0>0\repeat}}
\def\tableau#1{\vcenter{\offinterlineskip
  \tableaustep=\tableauside\advance\tableaustep by-\tableaurule
  \kern\normallineskip\hbox
    {\kern\normallineskip\vbox
      {\gettableau#1 0 }%
     \kern\normallineskip\kern\tableaurule}%
  \kern\normallineskip\kern\tableaurule}}
\def\gettableau#1 {\ifnum#1=0\let\next=\null\else
  \squares{#1}\let\next=\gettableau\fi\next}

 \def\eqnn#1{\xdef #1{(\secsym\the\meqno)}\writedef{#1\leftbracket#1}%
 \global\advance\meqno by1\wrlabeL#1}
 \def\eqna#1{\xdef #1##1{\hbox{$(\secsym\the\meqno##1)$}}
 \writedef{#1\numbersign1\leftbracket#1{\numbersign1}}%
 \global\advance\meqno by1\wrlabeL{#1$\{\}$}}
 \def\eqn#1#2{\xdef #1{(\secsym\the\meqno)}\writedef{#1\leftbracket#1}%
 \global\advance\meqno by1$$#2\eqno#1\eqlabeL#1$$}

\global\newcount\itemno \global\itemno=0

\def\itemaut#1{\global\advance\itemno by1\noindent\item{\the\itemno.}#1}

\def\({\left(}
\def\){\right)}

\def\pp{{\bf p}}

\def\lsim{\mathrel{\mathstrut\smash{\ooalign{\raise2.5pt\hbox{$<$}\cr\lower2.5pt\hbox{$\sim$}}}}}
\def\gsim{\mathrel{\mathstrut\smash{\ooalign{\raise2.5pt\hbox{$>$}\cr\lower2.5pt\hbox{$\sim$}}}}}

\def\overleftrightarrow#1{\vbox{\ialign{##\crcr
     $\leftrightarrow$\crcr\noalign{\kern-0pt\nointerlineskip}
     $\hfil\displaystyle{#1}\hfil$\crcr}}}
     
     \def\overleftarrow#1{\vbox{\ialign{##\crcr
     $\leftarrow$\crcr\noalign{\kern-0pt\nointerlineskip}
     $\hfil\displaystyle{#1}\hfil$\crcr}}}

\usepackage[yyyymmdd,hhmmss]{datetime}
\newif{\ifeq}           
\eqtrue                 
\newcounter{lecturecounter}